\documentclass[journal]{IEEEtran}
\usepackage{ifpdf}

\usepackage{cite}

\ifCLASSINFOpdf
   \usepackage[pdftex]{graphicx}
\else
   \usepackage[dvips]{graphicx}
\fi
\usepackage{amsmath}
\usepackage{algorithmic}

\usepackage{array}
\usepackage{url}

\usepackage{graphicx}
\usepackage{amsfonts}
\usepackage{bm}
\usepackage{amssymb}
\usepackage{multirow}
\usepackage{algorithm}
\usepackage{mathtools}

\begin{document}
%
\title{Energy Harvesting in Secure MIMO Systems}
%
%
%

\author{Wei~Wu,~\IEEEmembership{Member,~IEEE,}
        Xueqi~Zhang, Baoyun~Wang,~\IEEEmembership{Member,~IEEE,}
\thanks{Wei Wu is with the College of Communication and Information, Nanjing University of Posts and Telecommunications,
Nanjing, 210003, China (e-mail: bywunjupt@foxmail.com).}
\thanks{X. Zhang and B. Wang are with the College of Communication and Information, Nanjing University of Posts and
Telecommunications, Nanjing, 210003, China (e-mail: zhangxueqi9231@163.com, bywang@njupt.edu.cn).}
\thanks{This paper was supported by the National Natural Science Foundation
of China (No. 61271232, 61372126, 61501171); the Open research fund of National Mobile Communications Research Laboratory,
Southeast University (No. 2012D05); the Priority Academic Program Development of Jiangsu Province (Smart Grid and Control
Technology) for funding.}}

\maketitle

\begin{abstract}
The problems of energy harvesting in wireless communication systems have recently drawn much attention. In this paper, we focus on the investigation of energy harvesting maximization (EHM) in the important secrecy multi-input multi-output (MIMO) systems where little research has been done due to their complexity. Particularly, this paper studies the resource allocation strategies in MIMO wiretap channels, wherein we attempt to maximize the harvested energy by one or multiple multi-antenna energy receivers (ERs) (potential eavesdropper) while guaranteeing the secure communication for the multi-antenna information receiver (IR). Two types of IR, with and without the capability to cancel the interference from energy signals, are taken into account. In the scenario of single energy receiver (ER), we consider the joint design of the transmission information and energy covariances for EHM. Both of the optimization problems for the two types of IR are non-convex, and appear to be difficult. To circumvent them, the combination of first order Taylor approximation and sequential convex optimization approach is proposed. Then, we extend our attention to the scenario with multiple ERs, where the artificial noise (AN) aided weighted sum-energy harvesting maximization (WS-EHM) problem is considered. Other than the approaches adopted in solving the EHM problems, an algorithm conducts in an alternating fashion is proposed to handle this problem. In particular, we first perform a judicious transformation of the WS-EHM problem. Then, a block Gauss-Seidel (GS) algorithm based on logarithmic barrier method and gradient projection (GP) is derived to obtain the optimal solution of the reformulation by solving convex problems alternately. Furthermore, the resulting block GS method is proven to converge to a Karush-Kuhn-Tucker (KKT) point of the original WS-EHM problem. Simulation results show the outstanding energy harvesting performance and computational efficiency of our proposed algorithms.
\end{abstract}


%
\IEEEpeerreviewmaketitle

\section{Introduction}
%
%
%
%
\IEEEPARstart{W}{ith} the rapid development of wireless communication technology and the incremental user requirement, the ubiquitous high data rate becomes a basic merit of our next generation wireless communication networks, especially in the future 5G system. However, it indeed implicits a rapidly escalated energy consumption of the communication nodes serving in their corresponding networks to maintain such a property. Consequently, the green communication technologies for the purpose of decreasing the energy dissipated by each unit bit information transmission as much as possible attract much attention among researchers \cite{IEEEhowto:1, IEEEhowto:2, IEEEhowto:3, IEEEhowto:4, IEEEhowto:5, IEEEhowto:6, IEEEhowto:7, IEEEhowto:8, IEEEhowto:9}. Although the energy efficiency appears to be enhanced significantly by bringing in these advanced technologies, the limited lifetime of the wireless networks still remains as a bottleneck of its own development since the mobile devices are powered by batteries with limited action time. In general, one may recharge or replace the batteries to prolong the lifetime of the mobile devices. Nevertheless, it would lead to not cost-effective, inconvenient (say, in harsh environments or transducers embedded in human bodies or buildings) or environmentally-unfriendly. Hence, a cost-efficient, convenient, as well as eco-friendly alternative which can assist in capturing energy from surroundings is proposed. Other than the traditional sources such as solar, water conservancy, and biomass, etc., the ambient radio frequency (RF) signals turn out to be a feasible new source used to scavenge energy.

Obviously, it is a promising alternative to incorporate energy harvesting into wireless communication devices, which not only provides the perpetual renewable energy but also it is particularly suitable for low-power communication nodes. Many researchers from all over the world paid much attention to the RF-based energy scavenging systems in recent decades. The original recorded study of this field can be traced back to the 1914’s in Nikola Tesla's work \cite{IEEEhowto:1}, where some preliminary attempts of wireless power transfer about sensor networks, body area networks and passive RF identification systems are demonstrated. On the other hand, in accordance with recent research literatures \cite{IEEEhowto:3, IEEEhowto:4, IEEEhowto:5, IEEEhowto:6, IEEEhowto:7, IEEEhowto:8, IEEEhowto:9}, the RF-based energy transfer technology can facilitate simultaneous wireless information and power transfer (SWIPT) since the signals used to carry energy are able to serve as a vehicle for message transportation. Such a discovery brings numerous new challenges and opportunities to the engineers in this field. Authors in \cite{IEEEhowto:3} first came up with a capacity-energy function and studied the essential tradeoffs between information decoding rate and power transfer for the single-input single-output (SISO) systems. Literature \cite{IEEEhowto:4} extended the situation in \cite{IEEEhowto:3} to a frequency selection single antenna additive white Gaussian noise (AWGN) channels, and proposed an efficient resource allocation algorithm for the near-field communication systems they concerned. The works in \cite{IEEEhowto:5, IEEEhowto:6, IEEEhowto:7} further pursued the multi-input single-output (MISO) systems with multi-antenna transmitters. By exploiting the available degrees of freedom offered by multi-antenna transmitter, the energy transfer efficiency can be improved distinctly. The common three nodes MIMO broadcast channels were investigated in \cite{IEEEhowto:8} and \cite{IEEEhowto:9}. As a compromise, Zhang et al. \cite{IEEEhowto:8} presented two practical schemes, i.e., time switching and power splitting, for SWIPT. In \cite{IEEEhowto:9}, a transmission covariance optimization strategy was studied for the EHM problem.

From the papers outlined above, a consensus was revealed that more transmit power will be consumed for the purpose of higher energy scavenging at ER. However, in fact, an increasing transmission power means a higher susceptibility to eavesdropping due to the result of a substantial information leakage and the broadcasting nature of channels. Therefore, how communication security can be ensured arises as a crucial issue in SWIPT systems.

Recently, physical-layer security draws much attention in wireless communications. Different from the traditional way employing cryptographic encryption at the application layer, physical-layer security provides perfectly security based on the knowledge of physical properties of wireless channels. In the 1970’s, Wyner who first studied physical-layer security and proved a positive secrecy rate for degrade channels in \cite{IEEEhowto:10}. Notable extension of Wyner’s work included in \cite{IEEEhowto:11} and \cite{IEEEhowto:12}, where the non-degraded channels and the SISO Gaussian wiretap channels are considered, respectively. Furthermore, from the perspective of information theory, Loyka et al. \cite{IEEEhowto:13} obtained the optimal transmit covariance matrix over general three-node MIMO secrecy channels (not just the special channels mentioned in \cite{IEEEhowto:11} and \cite{IEEEhowto:12}) and gave a direct proof of necessary condition of optimality based on the necessary Karush-Kuhn-Tucker (KKT) conditions. A custom-designed algorithm taking advantage of extended barrier method was proposed in \cite{IEEEhowto:14}. Li et al. \cite{IEEEhowto:15} designed an AN-aided resource allocation strategy to maximize the achievable secrecy rate for an MISO channel since AN has been found to be a promising tool for preventing eavesdropping. Then, \cite{IEEEhowto:16} extended their works to the MIMO case, where an alternating optimization (AO) method to handle the secrecy optimization problems was derived through a judicious SCM (secrecy capacity maximization) reformulation. Wang et al. \cite{IEEEhowto:18} considered a two-hop AF relay network, a generalized singular value decomposition (SVD) based precoding approach was proposed to acquire the optimal power allocation for secrecy capacity. Besides, it is also note that the secure communication in SWIPT has been widely studied in the literatures \cite{IEEEhowto:20, IEEEhowto:21, IEEEhowto:22, IEEEhowto:23, IEEEhowto:24, IEEEhowto:25, IEEEhowto:26, IEEEhowto:27}, where the trade-off designs between secrecy rate and energy scavenging are achieved in different situations.

However, it is worth to note that the aforementioned works focusing on energy harvesting while meeting the secure communication requirement at IR are limited to the MISO systems or even the SISO systems. The more complicate MIMO systems still remain less studied in this field (most are confined to secrecy rate or energy efficiency maximization), though a preliminary work has been done in \cite{IEEEhowto:9}. In this paper, we consider the energy harvesting problems for MIMO secrecy systems with one or multiple multi-antenna ERs. Our problem formulations contain two types of IR, namely, the IR with and without capability to perform energy signals interference cancellation. The main contributions of this paper are summarized as follows.

\begin{itemize}
  \item For the one ER case (a common three node communication system), two EHM problems with the optimization of transmit covariance matrixes are considered. The formulated optimization problems, denoted by (P1) and (P2), are non-convex and subject to secrecy rate constraint for IR and given total power budget of the transmitter. By resorting to the Taylor series expansion, we reformulate both problems into convex ones. After that, for (P1), we further convert it into a two-layer optimization problem. Then, the iterative WF-SVD (water filling and SVD) algorithm resulting in the semi-closed form solutions and the golden section search algorithm are proposed for the inner layer and outer layer optimizations, resp.. For (P2), the off-the-shelf convex optimization tool, such as SeDuMi \cite{IEEEhowto:28} (also called as CVX \cite{IEEEhowto:29}), is utilized to obtain its optimal solution.
  \item For the multiple ERs case, the AN-aided WS-EHM problem is considered. And, different from above case, since the distinction caused by two types IR turn out to be obscure from the perspective of optimization, only one problem is studied. The problem formulation is non-convex and hard to tackle. To conquer this difficulty, a judicious reformulation for such an optimization framework is produced, which leads to the development of an efficient block GS \cite{IEEEhowto:30} algorithm. Particularly, first, we transform the original optimization problem into another equivalent problem by introducing several auxiliary variables. Next, based on the concept of AO, we propose a custom-made block GS algorithm to optimize the coupled variables, in which both of the gradient projection (GP) method \cite{IEEEhowto:31} and barrier method \cite{IEEEhowto:32} are incorporated and available for the construction of an algorithm with high computational efficiency. Finally, we prove the convergence of our proposed algorithm by guaranteeing the KKT \cite{IEEEhowto:32} optimality under the MIMO multiple ERs scenario.
\end{itemize}

The rest of this paper is organized as follows. Section II describes the system model and problem statement. Section III and Section IV provide the main contributions of this paper, wherein the efficient resource allocation algorithms are developed for the EHM and WS-EHM problems, respectively. In Section V, we illustrate numerical results to assess the performance of our proposed algorithms. A conclusion is made in Section VI.

\textit{Notations:} The lower-case, boldface lower-case and boldface upper-case letters are used to denote scalars, vectors and matrices, respectively; ${\rm{Tr}}\left( {\bf{X}} \right)$, $\left| {\bf{X}} \right| \left( {\det ({\bf{X}})} \right)$, ${{\bf{X}}^H}$, ${\left( {\bf{X}} \right)^{ - 1}}$,  ${\left\| {\bf{X}} \right\|_F}$ and ${\rm{rank}}\left( {\bf{X}} \right)$ denote trace, determinant, Hermitian transpose, inverse, Frobenius norm and rank of a matrix ${\bf{X}}$; ${\bf{I}}$ represents the identity matrix; ${\mathbb{C}^{n \times m}}$ denotes the space of $n \times m$ complex matrices; ${\bf{x}} \sim \mathcal{CN}\left( {{\bf{\mu }}, {\bf{\Sigma }}} \right)$ means that the random vector ${\bf{x}}$ follows a circularly symmetric complex Gaussian (CSCG) distribution with mean ${\bf{\mu }}$ and covariance ${\bf{\Sigma }}$; ${\bf{X}} \succeq {\bf{0}} \left( {{\bf{X}} \succ {\bf{0}}} \right)$ means that ${\bf{X}}$ is positive semidefinite (definite); ${\bf{S}}_{ +  + }^n$ denotes the symmetric positive definite matrices while ${\mathbb{H}}_ + ^n$ represents $n \times n$ Hermitian positive semidefinite matrices; ${\mathbb{R}^n}$ means n-dimensional real vectors and ${\mathbb{R}_{ +  + }}$ denotes the set of all positive real numbers.

\section{System Model and Problem Statement}
\subsection{System Model}

\begin{figure}
  \centering
  \includegraphics[width=3.0in]{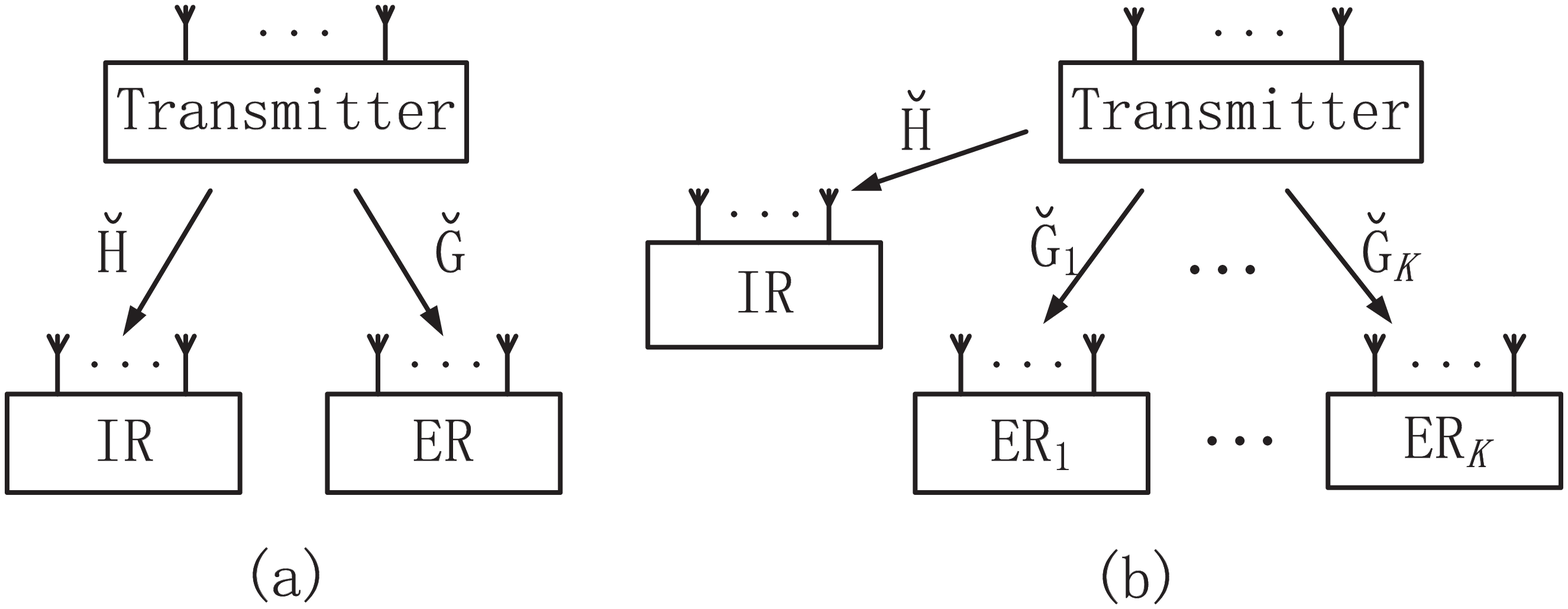}
  \caption{System model. (a) One multi-antenna ER. (b) Multiple multi-antenna ERs.}\label{fig:system}
\end{figure}

Consider a standard three-node MIMO downlink system as shown in Fig. \ref{fig:system}(a), which consists of a transmitter, one IR and one ER. The transmitter who wants to have a secure communication with IR and transmit energy to ER is equipped with ${N_t}$ antennas, while the IR and ER are equipped with ${N_i}$ and ${N_e}$ antennas, resp.. We assume that the ER also acts as a potential eavesdropper to intercept the confidential information intended for IR since the ER is located in the range of service coverage. The channel coefficients from the transmitter to IR and ER are denoted as  ${\bf{\mathord{\buildrel{\lower3pt\hbox{$\scriptscriptstyle\smile$}}\over H} }} \in {\mathbb{C}^{{N_t} \times {N_i}}}$ and ${\bf{\mathord{\buildrel{\lower3pt\hbox{$\scriptscriptstyle\smile$}}\over G} }} \in {\mathbb{C}^{{N_t} \times {N_e}}}$, resp.. Therefore, the received signals at IR and ER can be expressed as
\begin{equation}\label{eq:F1}
\begin{array}{l}
{{\bf{y}}_i} = {{{\bf{\mathord{\buildrel{\lower3pt\hbox{$\scriptscriptstyle\smile$}}
\over H} }}}^H}{\bf{x}} + {{\bf{n}}_i},\\
{{\bf{y}}_e} = {{{\bf{\mathord{\buildrel{\lower3pt\hbox{$\scriptscriptstyle\smile$}}
\over G} }}}^H}{\bf{x}} + {{\bf{n}}_e},
\end{array}
\end{equation}
resp., where ${\bf{x}} \in {\mathbb{C}^{{N_t} \times 1}}$ is the transmit symbol vector, ${{\bf{n}}_i} \sim \mathcal{CN}\left( {{\bf{0}}, \sigma _i^2{\bf{I}}} \right)$ and ${{\bf{n}}_e} \sim \mathcal{CN}\left( {{\bf{0}}, \sigma _e^2{\bf{I}}} \right)$ are i.i.d complex additive white Gaussian noises stem from the receiving antennas at the desired receiver. To facilitate SWIPT, the transmit symbol vector is selected as
\begin{equation}\label{eq:F2}
{\bf{x}} = {{\bf{w}}_i} + {{\bf{w}}_e},
\end{equation}
where ${{\bf{w}}_i} \in {\mathbb{C}^{{N_t} \times 1}}$ and ${{\bf{w}}_e} \in {\mathbb{C}^{{N_t} \times 1}}$ are the coded confidential information vector intended for IR and the Gaussian pseudo-random sequences to promote energy transfer, resp.. And ${{\bf{w}}_i}$ is assumed to comply with a complex Gaussian distribution ${{\bf{w}}_i} \sim \mathcal{CN}\left( {{\bf{0}}, {{\bf{W}}_I}} \right)$, where ${{\bf{W}}_I} \succeq {\bf{0}}$ is the transmit information covariance; ${{\bf{w}}_e}$ is modeled as a complex Gaussian pseudo-random vector which is distributed as ${{\bf{w}}_e} \sim \mathcal{CN}\left( {{\bf{0}}, {{\bf{W}}_E}} \right)$, where ${{\bf{W}}_E} \succeq {\bf{0}}$ is the covariance of energy signals. Both of the information and energy signals covariance matrices are the variables to be optimized in our following problems.

\subsection{Problem Statement}

Here, we focus on the EHM problems through the joint design of transmit information and energy covariances under the constraints of quality of service (QoS) requirement on secure communication and given total transmit power budget. It is worth pointing out that the pseudo-random signals which are used for delivering energy are always known at both the transmitter and IR before transmission, hence, the consequent interference can be cancelled at IR as long as the relevant operation is carried out. Accordingly, we consider two types of IRs with and without the ability to execute the interference cancellation about energy signals, and each type corresponds to an EHM problem.

Due to the broadcast nature of wireless communications, assuming unit slot duration, the total energy captured by ER from the received signals ${{\bf{y}}_e}$ can be given by \cite{IEEEhowto:5, IEEEhowto:6, IEEEhowto:7, IEEEhowto:8}:
\begin{equation}\label{eq:F3}
{\rm{E}}\left( {{{\bf{W}}_I}, {{\bf{W}}_E}} \right) = \eta {\rm{Tr}}\left[ {{{{\bf{\mathord{\buildrel{\lower3pt\hbox{$\scriptscriptstyle\smile$}}
\over G} }}}^H}\left( {{{\bf{W}}_I} + {{\bf{W}}_E}} \right){\bf{\mathord{\buildrel{\lower3pt\hbox{$\scriptscriptstyle\smile$}}
\over G} }}} \right],
\end{equation}
where $0 < \eta  \le 1$, a constant, is the energy conversion efficiency.

On the other hand, for the purpose of secure communication, given $\left( {{{\bf{W}}_I}, {{\bf{W}}_E}} \right)$, the achievable secrecy capacity at IR can be depicted as \cite{IEEEhowto:33}:
\begin{equation}\label{eq:F4}
{C_s} = {C_I}\left( {{{\bf{W}}_I}, {{\bf{W}}_E}} \right) - {C_E}\left( {{{\bf{W}}_I}, {{\bf{W}}_E}} \right),
\end{equation}
where ${C_I}\left( {{{\bf{W}}_I}, {{\bf{W}}_E}} \right)$ is the channel capacity between the transmitter and IR while ${C_E}\left( {{{\bf{W}}_I}, {{\bf{W}}_E}} \right)$ is the channel capacity between the transmitter and ER. From (\ref{eq:F4}), we can conclude that a perfect secrecy can be ensured when the IR decodes the confidential messages at ${C_s}$ bits per channel usage. In other words, at such a decoding rate, the potential eavesdropper (i.e., ER) can retrieve nearly nothing from the messages intended for IR.

In the following context, two EHM problems will be studied for two types of IRs. First, we focus on the case that the IR possesses the capability to cancel the interference from energy signals. On the other hand, of course, interference cancellation can also be performed at the ER when it serves as the potential eavesdropper \cite{IEEEhowto:6}. The channel capacities ${C_{I1}}\left( {{{\bf{W}}_I}, {{\bf{W}}_E}} \right)$ and ${C_{E1}}\left( {{{\bf{W}}_I}, {{\bf{W}}_E}} \right)$ can be written as:
\begin{equation}\label{eq:F5}
\begin{array}{l}
{C_{I1}}\left( {{{\bf{W}}_I}, {{\bf{W}}_E}} \right) = {\log}\left| {{\bf{I}} + \frac{1}{{\sigma _i^2}}{{{\bf{\mathord{\buildrel{\lower3pt\hbox{$\scriptscriptstyle\smile$}}
\over H} }}}^H}{{\bf{W}}_I}{\bf{\mathord{\buildrel{\lower3pt\hbox{$\scriptscriptstyle\smile$}}
\over H} }}} \right|,\\
{C_{E1}}\left( {{{\bf{W}}_I}, {{\bf{W}}_E}} \right) = {\log}\left| {{\bf{I}} + \frac{1}{{\sigma _e^2}}{{{\bf{\mathord{\buildrel{\lower3pt\hbox{$\scriptscriptstyle\smile$}}
\over G} }}}^H}{{\bf{W}}_I}{\bf{\mathord{\buildrel{\lower3pt\hbox{$\scriptscriptstyle\smile$}}
\over G} }}} \right|,
\end{array}
\end{equation}
respectively. For notational simplicity, we set ${\bf{H}} = \frac{1}{{{\sigma_i}}}{\bf{\mathord{\buildrel{\lower3pt\hbox{$\scriptscriptstyle\smile$}}
\over H} }}$ and ${\bf{G}} = \frac{1}{{{\sigma _e}}}{\bf{\mathord{\buildrel{\lower3pt\hbox{$\scriptscriptstyle\smile$}}
\over G} }}$, which are available in the following expression. Then, the first EHM design problem can be formulated as:
\begin{subequations}\label{eq:F6}
\begin{equation}\label{eq:F6a}
(\textsc{P1}): ~~\mathop {\max }\limits_{{{\bf{W}}_I} \succeq {\bf{0}},{\kern 1pt} {\kern 1pt} {{\bf{W}}_E} \succeq {\bf{0}}} {\kern 1pt} {\kern 1pt} {\kern 1pt} {\kern 1pt} \sigma _e^2\eta {\rm{Tr}}\left[ {{{\bf{G}}^H}\left( {{{\bf{W}}_I} + {{\bf{W}}_E}} \right){\bf{G}}} \right]
\end{equation}
\begin{equation}\label{eq:F6b}
 s.t. ~~ \underbrace {{{\log }}\left| {{\bf{I}} + {{\bf{H}}^H}{{\bf{W}}_I}{\bf{H}}} \right| - {{\log }}\left| {{\bf{I}} + {{\bf{G}}^H}{{\bf{W}}_I}{\bf{G}}} \right|}_{{C_{s1}}} \ge {C_0},
\end{equation}
\begin{equation}\label{eq:F6c}
 \rm{Tr}\left( {{{\bf{W}}_I} + {{\bf{W}}_E}} \right) \le \emph{P},
\end{equation}
\end{subequations}
where ${C_0} > 0$ is the target secrecy capacity for IR and $P > 0$ specifies the total power budget\footnote[1]{The constraint (\ref{eq:F6b}) turns out to be concave and the left side of this inequality is positive when the channel is degraded, i.e., ${\bf{H}}{{\bf{H}}^H} - {\bf{G}}{{\bf{G}}^H} \succeq {\bf{0}}$ holds \cite{IEEEhowto:9}.}.

As mentioned in the introduction, a simplified version of problem (P1) wherein only one optimization variable ${{\bf{W}}_I}$ is involved had ever been studied in \cite{IEEEhowto:9}. However, the solution proposed in \cite{IEEEhowto:9} is no longer appropriate for problem (P1) because of the additional variable ${{\bf{W}}_E}$ which leads to a more complicate problem in terms of optimization. In the next section, we will propose a valid solution for problem (P1) by dividing it into several subproblems.

Then, we also interest in the EHM problem with the IR not possessing the ability to perform interference cancellation. In this case, the channel capacities ${C_{I2}}\left( {{{\bf{W}}_I}, {{\bf{W}}_E}} \right)$ and ${C_{E2}}\left( {{{\bf{W}}_I}, {{\bf{W}}_E}} \right)$ can be given by
\begin{subequations}\label{eq:F7}
\begin{equation}\label{eq:F7a}
{C_{I2}}\left( {{{\bf{W}}_I}, {{\bf{W}}_E}} \right) = {\log}\left| {{\bf{I}} + {{\left( {{\bf{I}} + {{\bf{H}}^H}{{\bf{W}}_E}{\bf{H}}} \right)}^{ - 1}}{{\bf{H}}^H}{{\bf{W}}_I}{\bf{H}}} \right|,
\end{equation}
\begin{equation}\label{eq:F7b}
{C_{E2}}\left( {{{\bf{W}}_I}, {{\bf{W}}_E}} \right) = {\log}\left| {{\bf{I}} + {{\bf{G}}^H}{{\bf{W}}_I}{\bf{G}}} \right|,
\end{equation}
\end{subequations}
resp.. Therefore, the second EHM problem can be stated as:
\begin{subequations}\label{eq:F8}
\begin{equation}\label{eq:F8a}
(\textsc{P2}): ~~\mathop {\max }\limits_{{{\bf{W}}_I} \succeq {\bf{0}},{\kern 1pt} {\kern 1pt} {{\bf{W}}_E} \succeq {\bf{0}}} {\kern 1pt} {\kern 1pt} {\kern 1pt} {\kern 1pt} \sigma _e^2\eta {\rm{Tr}}\left[ {{{\bf{G}}^H}\left( {{{\bf{W}}_I} + {{\bf{W}}_E}} \right){\bf{G}}} \right]
\end{equation}
\begin{equation}\label{eq:F8b}
 s.t. ~~~ \underbrace {{C_{I2}}\left( {{{\bf{W}}_I}, {{\bf{W}}_E}} \right) - {C_{E2}}\left( {{{\bf{W}}_I}, {{\bf{W}}_E}} \right)}_{{C_{s2}}} \ge {C_0},
\end{equation}
\begin{equation}\label{eq:F8c}
 \rm{Tr}\left( {{{\bf{W}}_I} + {{\bf{W}}_E}} \right) \le \emph{P}.
\end{equation}
\end{subequations}
It can be seen that problem (P2) is a bit more complex than (P1) due to the modified constraint (\ref{eq:F8b}). The solution of problem (P2) will also be demonstrated in the following section.


\section{Resource Allocation Solutions for EHM Problems}

Owing to the fact that both the constraints (\ref{eq:F6b}) and (\ref{eq:F8b}) are non-concave functions with respect to (w.r.t.) ${{\bf{W}}_I}$ and ${{\bf{W}}_E}$, without doubt, the optimization problems (P1) and (P2) are both non-convex. Thus, the off-the-shelf tools or methods (such as SeDuMi \cite{IEEEhowto:28}, Lagrange duality \cite{IEEEhowto:32}, ect.) used to deal with convex optimization problems do not apply for them. To solve and derive the valid resource allocation strategies for the EHM problems (P1) and (P2), in this section, we pursue the solutions to these two problems, resp..

\subsection{A Water Filling and SVD Based Approach to Problem (P1)}

From (\ref{eq:F6}), it is easy to know that the major difficulty in solving problem (P1) lies in the non-concave constraint (\ref{eq:F6b}) which appears to be a fashion of difference of two concave functions. In fact, such a Shannon capacity expression attracts much attention recently and some solutions used for tackling it have been obtained \cite{IEEEhowto:13}, \cite{IEEEhowto:14}, \cite{IEEEhowto:16}, \cite{IEEEhowto:24}. In \cite{IEEEhowto:13}, the authors assumed a degraded channel model and transmit signals over the positive direction of this channel model to obtain an optimal signaling strategy. Under the assumption, a concave secrecy capacity expression is achieved, and consequently, the corresponding solution. One can refer to \cite{IEEEhowto:24} for an extension work, where the approach is only limited to the problem concentrate on secrecy rate maximization (SRM). Authors in \cite{IEEEhowto:14} rewrite the secrecy capacity by a minimax representation through a channel enhancement argument and a clever bounding technique. Whereas, similar to \cite{IEEEhowto:24}, this method works if and only if the secrecy rate is treated as objective. Another common method \cite{IEEEhowto:16} transforms the secrecy rate expression to an equivalent form by introducing auxiliary variables.

Here, to coordinate with our follow-up further processing, different from aforementioned approaches, we resort to the first order Taylor approximation to reformulate the secrecy rate constraint (\ref{eq:F6b}). Do expansion at any feasible point ${{\bf{W}}_{I0}} \succeq {\bf{0}}$,\footnote[2]{It is worth to note that the implication ${\tilde C_{s1}} \ge {C_0} \Rightarrow {C_{s1}} \ge {C_0}$ holds since ${\tilde C_{s1}} \le {C_{s1}}$ fulfills at any point ${{\bf{W}}_I} \succeq {\bf{0}}$ (${\log}\left| {{\bf{I}} + {{\bf{G}}^H}{{\bf{W}}_I}{\bf{G}}} \right|$ is concave about ${{\bf{W}}_I}$), and the equality between ${\tilde C_{s1}}$ and ${C_{s1}}$ satisfies if and only if ${{\bf{W}}_{I0}} = {{\bf{W}}_I}$, where ${{\bf{W}}_{I0}}$ is the update of ${{\bf{W}}_{I}}$ acquired in the last iteration.} we have \cite{IEEEhowto:9}
\begin{equation}\label{eq:F9}
\begin{array}{l}
{C_{s1}} \simeq {\log}\left| {{\bf{I}} + {{\bf{H}}^H}{{\bf{W}}_I}{\bf{H}}} \right| - {\log}\left| {{\bf{I}} + {{\bf{G}}^H}{{\bf{W}}_{I0}}{\bf{G}}} \right|\\
   ~~~~~~~- {\rm{Tr}}\left[ {\frac{1}{{\ln 2}}{\bf{G}}{{\left( {{\bf{I}} + {{\bf{G}}^H}{{\bf{W}}_{I0}}{\bf{G}}} \right)}^{ - 1}}{{\bf{G}}^H}\left( {{\bf{W}}_{I} - {{\bf{W}}_{I0}}} \right)} \right]\\
   ~\;~~~\buildrel \Delta \over = {{\tilde C}_{s1}}.
\end{array}
\end{equation}

After the approximation process on secrecy rate ${C_{s1}}$, it can be seen that the reformulation ${\tilde C_{s1}}$ is now a concave function w.r.t. the variable ${{\bf{W}}_I}$. Based on this, the EHM problem (P1) can be re-expressed as:
\begin{equation}\label{eq:F10}
\begin{array}{l}
(\widetilde {\textsc{P1}}): ~~\mathop {\max }\limits_{{{\bf{W}}_I} \succeq {\bf{0}},{\kern 1pt} {\kern 1pt} {{\bf{W}}_E} \succeq {\bf{0}}}  \sigma _e^2\eta {\rm{Tr}}\left[ {{{\bf{G}}^H}\left( {{{\bf{W}}_I} + {{\bf{W}}_E}} \right){\bf{G}}} \right]\\
~~~~~~~~~~~~~ s.t.~~~~~~{\kern 1pt} {\kern 1pt} {\kern 1pt} {\kern 1pt} {\kern 1pt} {\kern 1pt} {\kern 1pt} {\kern 1pt} {\kern 1pt}    {\kern 1pt} {\kern 1pt} {{\tilde C}_{s1}} \ge {C_0},\\
 ~~~~~~~~~~~~~~~~~~~~{\rm{Tr}}\left( {{{\bf{W}}_I} + {{\bf{W}}_E}} \right) \le P.
\end{array}
\end{equation}
Obviously, the representation $\left( {\widetilde {\textsc{P1}}} \right)$ is a convex semidefinite programming (SDP) \cite{IEEEhowto:32} problem, with fixed ${{\bf{W}}_{I0}}$, which can be conveniently and effectively solved by the existing convex optimization tools (e.g. SeDuMi \cite{IEEEhowto:28} and CVX \cite{IEEEhowto:29}) in hand in a globally optimal manner. Or, instead, solve its Lagrange dual problem based on the alternative optimization of the original variables and Lagrangian multipliers. However, it is inefficient in terms of computational efficiency, dealing with many SDPs during the update of the initial value ${{\bf{W}}_{I0}}$, no matter which solution is adopted. Therefore, here, we propose a water filling (WF) and SVD based approach to achieve a fast implementation for problem $\left( {\widetilde {\textsc{P1}}} \right)$.

From the construction of problem $\left( {\widetilde {\textsc{P1}}} \right)$, it is easy to learn the independence between variables ${{\bf{W}}_I}$ and ${{\bf{W}}_E}$. Hence, by introducing an auxiliary variable $0 < \alpha  \le 1$, the optimization problem (\ref{eq:F10}) can be divided into three subproblems
\begin{subequations}\label{eq:F11}
\begin{equation}\label{eq:F11a}
\left( {\widetilde {\textsc{P1}} - 1} \right):~~   \mathop {\max }\limits_{{{\bf{W}}_I} \succeq {\bf{0}}} {\kern 1pt} {\kern 1pt} {\kern 1pt} {\kern 1pt} \sigma _e^2\eta {\rm{Tr}}\left( {{{\bf{G}}^H}{{\bf{W}}_I}{\bf{G}}} \right)
\end{equation}
\begin{equation}\label{eq:F11b}
s.t. ~~~~{{\tilde C}_{s1}} \ge {C_0},
\end{equation}
\begin{equation}\label{eq:F11c}
~~~~~~~~~~{\rm{Tr}}\left( {{{\bf{W}}_I}} \right) \le \alpha P,
\end{equation}
\end{subequations}
\begin{equation}\label{eq:F12}
\begin{array}{l}
\left( {\widetilde {\textsc{P1}} - 2} \right):  ~~    \mathop {\max }\limits_{{{\bf{W}}_E} \succeq {\bf{0}}} {\kern 1pt} {\kern 1pt} {\kern 1pt} {\kern 1pt} \sigma _e^2\eta {\rm{Tr}}\left( {{{\bf{G}}^H}{{\bf{W}}_E}{\bf{G}}} \right)\\
~~~~~~~~~~~s.t.~~~~{\rm{Tr}}\left( {{{\bf{W}}_E}} \right) \le \left( {1 - \alpha } \right)P,
\end{array}
\end{equation}
\begin{equation}\label{eq:F13}
\begin{array}{l}
\left( {\widetilde {\textsc{P1}} - 3} \right): ~~ \mathop {\max }\limits_\alpha    h\left( \alpha  \right)\\
~~~~~~~~~~~s.t.  ~~~~  {\rm{0 < }}\alpha  \le {\rm{1,}}
\end{array}
\end{equation}
where $h\left( \alpha  \right)$ denotes the optimal value of problem $\left( {\widetilde {{\textsc{P1}}}} \right)$. In fact, above subproblems can be classified into a two-layer optimization problem, where the inner layer contains two convex problems $\left({\widetilde {\textsc{P1}} - 1}\right)$ and $\left({\widetilde {\textsc{P1}} - 2}\right)$ with fixed $\alpha $ while the outer layer is an one-dimensional search problem $\left({\widetilde {\textsc{P1}} - 3}\right)$ with variable $\alpha $.
\subsubsection{Inner Layer Optimization}

Two subproblems are included in the inner layer optimization. First, we focus on subproblem $\left({\widetilde {\textsc{P1}} - 1}\right)$. Our general idea to solve this subproblem is reformulating it into a mode that can be conveniently tackled by the WF method. Since problem $\left({\widetilde {\textsc{P1}} - 1}\right)$ is convex w.r.t. ${{\bf{W}}_I}$ and satisfies the Slater’s condition \cite{IEEEhowto:32}, it can be solved by exploiting its dual problem with zero duality gap. Thus, by introducing two non-negative dual variables, $\lambda $ and $\mu $, w.r.t. the secrecy capacity constraint and the transmit power constraint. We have the following Lagrangian and dual function:
\begin{subequations}\label{eq:F14}
\begin{equation}\label{eq:F14a}
\begin{array}{l}
{\mathcal{L}}\left( {\lambda , \mu , {{\bf{W}}_I}} \right) = \sigma _e^2\eta {\rm{Tr}}\left( {{{\bf{G}}^H}{{\bf{W}}_I}{\bf{G}}} \right) + \lambda \left( {{{\tilde C}_{s1}} - {C_0}} \right)\\
~~~~~~~~~~~~~~~~~~- \mu \left[ {{\rm{Tr}}\left( {{{\bf{W}}_I}} \right) - \alpha P} \right]
\end{array},
\end{equation}
\begin{equation}\label{eq:F14b}
d\left( {\lambda , \mu } \right) = \mathop {\max }\limits_{{{\bf{W}}_I} \succeq {\bf{0}}} {\kern 1pt} {\kern 1pt} {\mathcal{L}}\left( {\lambda , \mu , {{\bf{W}}_I}} \right).
\end{equation}
\end{subequations}
The dual problem of $\left({\widetilde {\textsc{P1}} - 1}\right)$, denoted as $\left({\widetilde {\textsc{P1}} - 1-D}\right)$, is defined as $\mathop {\min }\limits_{\lambda  > 0,{\kern 1pt} {\kern 1pt} \mu  > 0} {\kern 1pt} {\kern 1pt} d\left( {\lambda ,{\kern 1pt} {\kern 1pt} \mu } \right)$. To solve the dual problem $\left({\widetilde {\textsc{P1}} - 1-D}\right)$, first, we maximize the Lagrangian (\ref{eq:F14a}) about ${{\bf{W}}_I}$ with fixed $\lambda $ and $\mu $. Then, minimize (\ref{eq:F14b}) to obtain the optimal dual solutions ${\lambda ^*}$ and ${\mu ^*}$. Finally, the dual problem is resolved and the optimal primary solution ${\bf{W}}_I^*$ that maximizes (\ref{eq:F14a}) is obtained when $d\left( {{\lambda ^*}, {\mu ^*}} \right)$ is acquired, i.e., the complementary slackness conditions ${\lambda ^*}\left( {\tilde C_{_{s1}}^* - {C_0}} \right) = 0$ and ${\mu ^*}\left[ {{\rm{Tr}}\left( {{\bf{W}}_{_I}^*} \right) - \alpha P} \right] = 0$ hold \cite{IEEEhowto:32}, where $\tilde C_{_{s1}}^*$ is the optimal secrecy capacity. Moreover, the analytic form of the optimal solution ${\bf{W}}_I^*$ is worked out and relegated to the following theorem.
\newtheorem{Theorem}{\textit{\underline{Theorem}}}[section]
\begin{Theorem}\label{thm:1}
The optimal solution ${\bf{W}}_I^*$ of the subproblem $\left({\widetilde {\textsc{P1}} - 1}\right)$ can be formed as follows:
\begin{equation}\label{eq:F15}
{\bf{W}}_I^* = {{\bf{Q}}^{ - {1 \mathord{\left/
 {\vphantom {1 2}} \right.
 \kern-\nulldelimiterspace} 2}}}{\bf{V}}{{\bf{D}}^*}{{\bf{V}}^H}{{\bf{Q}}^{ - {1 \mathord{\left/
 {\vphantom {1 2}} \right.
 \kern-\nulldelimiterspace} 2}}},
\end{equation}
\end{Theorem}
where ${\bf{Q}} = {\lambda ^*}{\bf{G}}{\left( {{\bf{I}} + {{\bf{G}}^H}{{\bf{W}}_{I0}}{\bf{G}}} \right)^{ - 1}}{{\bf{G}}^H} - \sigma _e^2\eta {\bf{G}}{{\bf{G}}^H} + {\mu ^*}{\bf{I}} \succ {\bf{0}}$ (the implication ${\bf{Q}} \succ {\bf{0}}$ must be hold, one can refer to \cite{IEEEhowto:9} for details), ${\bf{V}} \in {\mathbb{C}^{{N_t} \times L}}$ is derived from the SVD of ${{\bf{H}}^H}{{\bf{Q}}^{ - {1 \mathord{\left/{\vphantom {1 2}} \right. \kern-\nulldelimiterspace} 2}}} = {\bf{U}}{{\bf{\Lambda }}^{{1 \mathord{\left/{\vphantom {1 2}} \right. \kern-\nulldelimiterspace} 2}}}{{\bf{V}}^H}$, with ${{\bf{\Lambda }}^{{1 \mathord{\left/{\vphantom {1 2}} \right.
 \kern-\nulldelimiterspace} 2}}} = {\rm{diag}}\left( {{\delta _1},{\kern 1pt} {\kern 1pt} {\kern 1pt} {\delta _2}, \cdots ,{\kern 1pt} {\kern 1pt} {\kern 1pt} {\delta _L}} \right){\kern 1pt} $, ${\delta _1} \ge {\delta _2} \ge  \cdots  \ge {\delta _L} > 0$, and ${{\bf{D}}^*} = {\rm{diag}}\left( {p_{_1}^*,{\kern 1pt} {\kern 1pt} {\kern 1pt} {\kern 1pt} {\kern 1pt} {\kern 1pt}  \cdots ,{\kern 1pt} {\kern 1pt} {\kern 1pt} p_{_L}^*} \right)$ with
 $p_{_i}^* = \left( {{\lambda ^*} - \frac{1}{{\delta _i^2}}} \right),{\kern 1pt} {\kern 1pt} {\kern 1pt} {\kern 1pt} i = 1,{\kern 1pt} {\kern 1pt} {\kern 1pt}  \cdots ,{\kern 1pt} {\kern 1pt} {\kern 1pt} L$
 obtained from the principle of WF power allocation.

\textit{Proof:} Please refer to Appendix A for details.

As mentioned in Appendix A, the ellipsoid method \cite{IEEEhowto:35} is utilized to optimize the Lagrangian multipliers $\lambda $ and $\mu $. In regarding to the update of these two dual variables, we have the following conclusions.
\newtheorem{Proposition}{\textit{\underline{Proposition}}}[section]
\begin{Proposition}\label{pro:1}
In the process of each iteration of $\lambda$ and $\mu$, we have ${\left[ {{\lambda ^j},{\kern 1pt} {\kern 1pt} {\kern 1pt} {\kern 1pt} {\mu ^j}} \right]^T} = {\left[ {{\lambda ^{j - 1}},{\kern 1pt} {\kern 1pt} {\kern 1pt} {\kern 1pt} {\mu ^{j - 1}}} \right]^T} - \frac{1}{3}{{\bf{A}}^{j - 1}}{\bf{\tilde s}}$, where ${\lambda ^j}$ and ${\mu ^j}$ are the solutions result from the \textit{j}th iteration, ${{\bf{A}}^j} = \frac{4}{3}\left( {{{\bf{A}}^{j - 1}} - \frac{2}{3}{{\bf{A}}^{j - 1}}{\bf{\tilde s}}{{{\bf{\tilde s}}}^T}{{\bf{A}}^{j - 1}}} \right)$ with given initialization ${{\bf{A}}^0} \in {\bf{S}}_{ +  + }^2$ and ${\bf{\tilde s}} = \frac{1}{{\sqrt {{{\bf{s}}^T}{{\bf{A}}^{j - 1}}{\bf{s}}} }}{\bf{s}}$ with the subgradient ${\bf{s}} = {\left[ {\tilde C_{_{s1}}^* - {C_0}, {\kern 1pt} \alpha P - {\rm{Tr}}\left( {{\bf{W}}_{_I}^*} \right)} \right]^T}$; the iteration stopping criterion is $stop = {\left( {{{\bf{s}}^T}{{\bf{A}}^j}{\bf{s}}} \right)^{{1 \mathord{\left/
 {\vphantom {1 2}} \right. \kern-\nulldelimiterspace} 2}}} \le \varepsilon  \left( {\varepsilon  > 0} \right),$ and the iterations will be no more than $8\ln \left( {\frac{{r{l_s}}}{\varepsilon }} \right)$, where $r$ is associated with ${{\bf{A}}^0}$ while ${l_s}$ is associated with the subgradient ${\bf{s}}$.
\end{Proposition}

\textit{Proof:} See Appendix B.

Next, after the discussion about subproblem $\left({\widetilde {\textsc{P1}} - 1}\right)$, let's focus on the other subproblem $\left({\widetilde {\textsc{P1}} - 2}\right)$. Also, similar to the result of previous subproblem, a closed form solution can be derived based on the specific structure of $\left({\widetilde {\textsc{P1}} - 2}\right)$. Following a parallel mean, which is employed in structural analysis about convex optimization problem, as in \cite{IEEEhowto:8}, the optimal solution ${\bf{W}}_E^*$ can be presented as follows.

\newtheorem{thm}{\textit{\underline{Theorem}}}[section]
\begin{Theorem}\label{thm:2}
The optimal solution ${\bf{W}}_E^*$ of subproblem $\left({\widetilde {\textsc{P1}} - 2}\right)$ can be given as:
\[{\bf{W}}_E^{\rm{*}}{\rm{ = }}\left( {1 - \alpha } \right)P{{\bf{u}}_1}{\bf{u}}_1^H,\]
where ${{\bf{u}}_1} \in {\mathbb{C}^{M \times 1}}$ represents the first column of the unitary matrix ${{\bf{U}}_G} \in {\mathbb{C}^{{N_t} \times {N_t}}}$, and ${{\bf{U}}_G}$ is the result from the SVD of the channel matrix ${\bf{G}}$ given by ${\bf{G}}{\rm{ = }}{{\bf{U}}_G}{\bf{\Lambda }}_G^{{1 \mathord{\left/{\vphantom {1 2}} \right. \kern-\nulldelimiterspace} 2}}{\bf{V}}_G^H$.
\end{Theorem}

\textit{Proof:} The upshot in proving Theorem \ref{thm:2} lies in the SVD to both the optimization variable ${{\bf{W}}_E}$ and the channel matrix ${\bf{G}}$. The proof is not difficult and the details are omitted for brevity. One can refer to \cite{IEEEhowto:8} for a similar proof.

It is noted that the optimal covariance matrix of the transmit energy signal ${\bf{W}}_E^{\rm{*}}$ is a rank-one matrix and the maximum harvested energy contributed by the energy signals is given by ${\textsc{E}_{\max }} = \sigma _e^2\eta \left( {1 - \alpha } \right){g_1}P$, where ${g_1}$ is the biggest diagonal element w.r.t. to the diagonal matrix ${{\bf{\Lambda }}_G}$. Such a maximum harvested energy reached if and only if the transmit beamforming is performed on the energy signals, which aligns with the most powerful eigenmode of the matrix ${\bf{G}}{{\bf{G}}^H}$. On the basis of above-produced WF and SVD procedures for the inner layer optimization, an iterative WF-SVD algorithm is derived and the details are summarized in Algorithm \ref{alg:Algorithm 1}.

\begin{algorithm}[!t]    
\caption{The iterative WF-SVD algorithm for the inner layer optimization.}   
\begin{algorithmic}[1] \label{alg:Algorithm 1}     
\STATE \textbf{Initialize:} $j=1$, $m=1$, $\lambda  \ge 0$, $\mu  \ge 0$, ${\varepsilon _1} > 0, {\varepsilon _2} > 0$, ${{\bf{W}}_{I0}} \succeq {\bf{0}}$, ${\bf{W}}_I^0 = {{\bf{W}}_{I0}}$ and ${\bf{A}} = {r^2}{\bf{I}} \succ {\bf{0}}$, where ${\bf{I}}$ is a two-dimensional unit matrix;\\
\STATE\textbf{Repeat}
  \STATE ~~Set ${{\bf{A}}^0} = {\bf{A}}$, ${\lambda ^0} = \lambda $, ${\mu ^0} = \mu $;
  \STATE ~~\textbf{Repeat}
    \STATE ~~~~Compute ${\bf{W}}_I^*$ through Theorem \ref{thm:1};
    \STATE ~~~~Calculate the subgradient ${\bf{s}}$;
    \STATE ~~~~Update ${\lambda ^j}$, ${\mu ^j}$ and ${{\bf{A}}^j}$ according to Proposition \ref{pro:1};
    \STATE ~~~~$j=j+1$;
  \STATE ~~\textbf{Until} $stop \le {\varepsilon _1}$;
  \STATE ~~${{\bf{W}}_{I0}} = {\bf{W}}_I^*$, ${\bf{W}}_{^I}^m = {\bf{W}}_I^*$ and $\Delta {{\bf{W}}_I} = {\bf{W}}_I^m - {\bf{W}}_I^{m - 1}$;
  \STATE ~~$m=m+1$;
\STATE \textbf{Until} ${\left\| {\Delta {{\bf{W}}_I}} \right\|_2} \le {\varepsilon _2}$;
\STATE Compute ${\bf{W}}_E^*$ through Theorem \ref{thm:2};
\STATE \textbf{output:} ${\bf{W}}_I^*$ and ${\bf{W}}_E^*$.
\end{algorithmic}
\end{algorithm}

\textit{Feasibility assurance about the initial value ${{\bf{W}}_{I0}}$:} The common way to set ${{\bf{W}}_{I0}}$ is to initialize it as zeros or a diagonal matrix with the full transmit power uniformly distributed on each element. However, in consideration of the complicated constraints in (\ref{eq:F11}), the initialization may not be feasible to constraint (\ref{eq:F11b}) when different channel realizations are performed. This will lead to the poor convergence performance and infeasible simulation result of our proposed algorithm. Therefore, to circumvent this defect, the feasibility assurance procedure is necessary to be injected in the determination of the initial value ${{\bf{W}}_{I0}}$. Since the preset ${{\bf{W}}_{I0}}$ satisfies the constraint (\ref{eq:F11c}) with fixed $\alpha $, our main energy will concentrate on the feasibility check of constraint (\ref{eq:F11b}). To this end, we proceed as follows. Constructing a SCM problem extracted from subproblem $\left({\widetilde {\textsc{P1}} - 1}\right)$
\begin{equation}\label{eq:F16}
\begin{array}{l}
       ~~~~~\mathop {\max }\limits_{{{\bf{W}}_I} \succeq {\bf{0}}} {\kern 1pt} {\kern 1pt} {\kern 1pt} {\kern 1pt} {{\tilde C}_{s1}}\\
 s.t. ~~~~{\rm{Tr}}\left( {{{\bf{W}}_I}} \right) \le \alpha P.
\end{array}
\end{equation}
Given the initialization ${{\bf{W}}_{I0}}$, problem (\ref{eq:F16}) can be solved effectively through the interior point method \cite{IEEEhowto:32}, and the final optimal objective $\tilde C_{s1}^*$, which must be greater than ${C_0}$, can be achieved based on sequential convex optimizations. As a consequence, the corresponding optimal variable ${\bf{W}}_I^*$ can be selected as the feasible initial matrix. Alternatively, we can pick out a feasible initialization from the sequential convex operations once the secrecy capacity lower bound ${C_0}$ is reached. On the other hand, if the initial matrix ${{\bf{W}}_{I0}}$ fulfills the secrecy capacity and total transmit power constraints, no feasibility check procedures are required.

Since both the subproblems $\left({\widetilde {\textsc{P1}} - 1}\right)$ and $\left({\widetilde {\textsc{P1}} - 2}\right)$ are convex, following with the arguments raised in \cite{IEEEhowto:36}, we have the provable guarantee on the convergence of our proposed algorithm as follows:
\newtheorem{Pro}{\textit{\underline{Proposition}}}[section]
\begin{Proposition}\label{pro:2}
The iterative operations produce non-descending objective values of subproblem $\left({\widetilde {\textsc{P1}} - 1}\right)$, and it is guaranteed that the optimal solutions (${\bf{W}}_I^*$ and ${\bf{W}}_E^*$) generated by Algorithm \ref{alg:Algorithm 1} fulfill the KKT optimality conditions of subproblems $\left({\widetilde {\textsc{P1}} - 1}\right)$ and $\left({\widetilde {\textsc{P1}} - 2}\right)$, resp..
\end{Proposition}

\subsubsection{Outer Layer Optimization}

A single-variable optimization problem $\left({\widetilde {\textsc{P1}} - 3}\right)$ is considered in the outer layer part. Since the interval of the single variable $\alpha $ is bounded and smaller than one, one-dimensional search techniques can be applied to pursue the solution of $\left({\widetilde {\textsc{P1}} - 3}\right)$. Here, we resort to the efficient golden section search approach which is summarized in Algorithm \ref{alg:Algorithm 2}.
\begin{algorithm}[!t]    
\caption{The golden section search algorithm for the outer layer optimization.}   
\begin{algorithmic}[1] \label{alg:Algorithm 2}     
\STATE \textbf{Initialize:} $\alpha  \in \left( {0 1} \right]$, $a = {{\left( {\sqrt 5  - 1} \right)} \mathord{\left/
 {\vphantom {{\left( {\sqrt 5  - 1} \right)} 2}} \right.\kern-\nulldelimiterspace} 2}$, $b = 0$, $c = 1$, $\zeta  > 0$, ${\alpha _1} = \left( {1 - a} \right)c$ and ${\alpha _2} = ac$\\
\STATE  Compute $h\left( {{\alpha _1}} \right)$ and $h\left( {{\alpha _2}} \right)$ based on the results of inner layer optimization;
\STATE\textbf{Repeat}
  \STATE ~~if $h\left( {{\alpha _1}} \right) > h\left( {{\alpha _2}} \right)$
  \STATE ~~~~~$c = {\alpha _2}$, ${\alpha _2} = {\alpha _1}$, ${\alpha _1} = c - a\left( {c - b} \right),$
  \STATE ~~~~~compute $h\left( {{\alpha _1}} \right)$ as in step 2;
  \STATE ~~else
  \STATE ~~~~~$b = {\alpha _1}$, ${\alpha _1} = {\alpha _2}$, ${\alpha _2} = b + a\left( {c - b} \right)$,
  \STATE ~~~~~compute $h\left( {{\alpha _2}} \right)$ as in step 2;
\STATE \textbf{Until} $\left| {c - b} \right| \le \zeta $.
\end{algorithmic}
\end{algorithm}

So far, we can cope with problem $\left({\widetilde {\textsc{P1}}}\right)$ by performing a one-dimensional search during which a series of SDPs are tackled based on the custom-designed algorithm. As a result, the optimal solution $\left( {{\bf{W}}_I^*, {\bf{W}}_E^*, {\alpha ^*}} \right)$ will be obtained. Moreover, according to \cite{IEEEhowto:36}, ${\bf{W}}_I^*$ fulfills the KKT conditions of the original problem $\left( {\textsc{P1}} \right)$ though a Taylor linear approximation is implemented about variable ${{\bf{W}}_I}$.

\subsection{Sequential Convex Optimization to Problem (P2)}

It is easy to see that, like problem (P1), problem (P2) is also non-convex owing to the complex non-convex secrecy capacity constraint (\ref{eq:F8b}). However, in our experience, the above mentioned sequential convex optimization framework based on first-order Taylor series expansion is also available for the dispose of this problem. In the light of such an idea, we first rewrite the secrecy capacity expression ${C_{s2}}$ as:
\begin{equation}\label{eq:F17}
\begin{array}{l}
{C_{s2}} = {\log}\left| {{\bf{I}} + {{\bf{H}}^H}\left( {{{\bf{W}}_I}{\rm{ + }}{{\bf{W}}_E}} \right){\bf{H}}} \right| - {\log}\left| {{\bf{I}} + {{\bf{H}}^H}{{\bf{W}}_E}{\bf{H}}} \right|\\
 ~~~~~~~- {\log}\left| {{\bf{I}} + {{\bf{G}}^H}{{\bf{W}}_I}{\bf{G}}} \right|.
\end{array}
\end{equation}
Then, for any given ${{\bf{W}}_{I0}} \succeq {\bf{0}}$ and ${{\bf{W}}_{E0}} \succeq {\bf{0}}$, Taylor expand the last two terms on the right side of equation (\ref{eq:F17}), we have
\begin{equation}\label{eq:F18}
\begin{array}{l}
{{\tilde C}_{s2}} \buildrel \Delta \over = {\log}\left| {{\bf{I}} + {{\bf{H}}^H}({{\bf{W}}_I}{\rm{ + }}{{\bf{W}}_E}){\bf{H}}} \right| - {\log}\left| {{\bf{I}} + {{\bf{H}}^H}{{\bf{W}}_{E0}}{\bf{H}}} \right|\\
 ~~~~~~ - {\rm{Tr}}\left[ {\frac{1}{{\ln 2}}{\bf{H}}{{({\bf{I}} + {{\bf{H}}^H}{{\bf{W}}_{E0}}{\bf{H}})}^{ - 1}}{{\bf{H}}^H}({{\bf{W}}_E} - {{\bf{W}}_{E0}})} \right]{\kern 1pt}  \\
  ~~~~~~- {\log}\left| {{\bf{I}} + {{\bf{G}}^H}{{\bf{W}}_{I0}}{\bf{G}}} \right|\\
 ~~~~~~- {\rm{Tr}}\left[ {\frac{1}{{\ln 2}}{\bf{G}}{{\left( {{\bf{I}} + {{\bf{G}}^H}{{\bf{W}}_{I0}}{\bf{G}}} \right)}^{ - 1}}{{\bf{G}}^H}\left( {{{\bf{W}}_I} - {{\bf{W}}_{I0}}} \right)} \right].
\end{array}
\end{equation}
Based on this approximation, the EHM problem, given any feasible initialization $\left( {{{\bf{W}}_{I0}}, {{\bf{W}}_{E0}}} \right)$, can be reformulated as:
\begin{equation}\label{eq:F19}
\begin{array}{l}
(\widetilde {\textsc{P2}}):~ \mathop {\max }\limits_{{{\bf{W}}_I} \succeq {\bf{0}},{\kern 1pt} {\kern 1pt} {{\bf{W}}_E} \succeq {\bf{0}}} {\kern 1pt} {\kern 1pt} {\kern 1pt} {\kern 1pt} \sigma _e^2\eta {\rm{Tr}}\left[ {{{\bf{G}}^H}\left( {{{\bf{W}}_I} + {{\bf{W}}_E}} \right){\bf{G}}} \right]\\
 ~~~~~~~~~~~~s.t.~~~~~{\kern 1pt} {\kern 1pt} {\kern 1pt} {\kern 1pt} {\kern 1pt} {\kern 1pt} {\kern 1pt} {\kern 1pt} {\kern 1pt} {\kern 1pt} {\kern 1pt}  {{\tilde C}_{s2}} \ge {C_0},\\
 ~~~~~~~~~~~~~~~~~~~{\rm{Tr}}\left( {{{\bf{W}}_I} + {{\bf{W}}_E}} \right) \le P.
\end{array}
\end{equation}
It is obvious that the reformulation is convex w.r.t. its variables ${{\bf{W}}_I}$ and ${{\bf{W}}_E}$, so it can be globally optimized by the state of the art interior point method with polynomial complexity. Then, in the light of the demonstration in [36], the KKT point, i.e., the optimal solution $\left( {{\bf{W}}_{_I}^*, {\bf{W}}_{_E}^*} \right)$, of the original problem (P2) can be determined by performing sequential convex optimization on problem $(\widetilde {\textsc{P2}})$.

\section{Resource Allocation for the AN-aided WS-EHM Problem}

In this section, we extend our attention to a generalization of the previously investigated three-node MIMO wiretap channels, where multiple ERs are present and the transmitter is allowed to generate AN \cite{IEEEhowto:5},\cite{IEEEhowto:6} for the purpose of promoting efficient energy transfer and guaranteeing secure communication. On this occasion, the WS-EHM problem subject to the constraints of given secrecy rate for IR and limited transmit power of transmitter is considered. Similar to the scenarios assumed in Section II, in this section, we also presume the presence of two types of IRs with and without ability to eliminate the interference incurred by energy signals. Nevertheless, other than the manner adopted in Section III (deal with each case on its merits), here, we only study one of the two scenarios and custom-design an efficient algorithm for the corresponding optimization problem while the other one is regarded as a simplified version of this one from the perspective of optimization. In order to tackle the resulting optimization problem, many mathematical techniques (such as barrier method, GP method and block GS method) are employed which will be elaborated in the following subsections.

\subsection{Signal Model and Problem Formulation}

In the secure MIMO multiple ERs communication system shown in Fig. \ref{fig:system}(b), the received signals at the IR and \textit{k}th ER can be given by
\begin{equation}\label{eq:F20}
\begin{array}{l}
{{\bf{y}}_i} = {{{\bf{\mathord{\buildrel{\lower3pt\hbox{$\scriptscriptstyle\smile$}}
\over H} }}}^H}{\bf{x}} + {{\bf{n}}_i},\\
{{\bf{y}}_{e,k}} = {\bf{\mathord{\buildrel{\lower3pt\hbox{$\scriptscriptstyle\smile$}}
\over G} }}_k^H{\bf{x}} + {{\bf{n}}_{e,k}}, k \in {\rm K},
\end{array}
\end{equation}
resp., where ${{\bf{n}}_{e,k}} \sim \mathcal{CN}\left( {{\bf{0}}, \sigma _{e,k}^2{\bf{I}}} \right)$ is the complex additive white Gaussian noise and ${\rm K} \buildrel \Delta \over = \left\{ {1,  \cdots , K} \right\}$. The transmit signal ${\bf{x}} \in {\mathbb{C}^{{N_t} \times 1}}$ is modeled as:
\[{\bf{x}} = {{\bf{w}}_i} + {{\bf{w}}_e} + {\bf{v}},\]
where the newly introduced vector ${\bf{v}} \in {\mathbb{C}^{{N_t} \times 1}}$ is the AN which is generated by the transmitter to facilitate both valid energy transfer and secure communication. The signal vectors ${{\bf{w}}_i}$ and ${{\bf{w}}_e}$ are assumed to follow the distribution ${{\bf{w}}_i} \sim \mathcal{CN}\left( {{\bf{0}}, {{\bf{W}}_I}} \right)$ and ${{\bf{w}}_e} \sim \mathcal{CN}\left( {{\bf{0}}, {{\bf{W}}_E}} \right)$, resp., as before, while ${\bf{v}}$ is modeled as a complex Gaussian vector with ${\bf{v}} \sim \mathcal{CN}\left( {{\bf{0}}, {\bf{V}}} \right)$.

Based on the setup above, considering the case that the IR is unable to eliminate the interference from energy signals, set ${\bf{H}} = \frac{1}{{{\sigma _i}}}{\bf{\mathord{\buildrel{\lower3pt\hbox{$\scriptscriptstyle\smile$}}
\over H} }}$ and ${{\bf{G}}_k} = \frac{1}{{{\sigma _{e,k}}}}{{\bf{\mathord{\buildrel{\lower3pt\hbox{$\scriptscriptstyle\smile$}}
\over G} }}_k}, \forall k \in {\rm K}$, we formulate the AN aided WS-EHM problem as follows:
\begin{subequations}\label{eq:F21}
\begin{equation}\label{eq:F21a}
({\textsc{P3}}) :   ~~    \mathop {\max }\limits_{{{\bf{W}}_I} \succeq {\bf{0}}, {{\bf{W}}_E} \succeq {\bf{0}}, {\bf{V}} \succeq {\bf{0}}}  {{\rm{E}}_{{\rm{WS}}}}\left( {{{\bf{W}}_I}, {{\bf{W}}_E}, {\bf{V}}} \right)
\end{equation}
\begin{equation}\label{eq:F21b}
s.t. ~~~~~~ {R_s} \ge {R_0}\ln 2,
\end{equation}
\begin{equation}\label{eq:F21c}
~~~~~~~~~~~{\rm{Tr}}\left( {{{\bf{W}}_I} + {{\bf{W}}_E} + {\bf{V}}} \right) \le P,
\end{equation}
\end{subequations}
where ${{\rm{E}}_{{\rm{WS}}}}\left( {{{\bf{W}}_I}, {{\bf{W}}_E}, {\bf{V}}} \right) = \sum\limits_{k = 1}^K {\sigma _{e,k}^2} {\mu _k}{\eta _k}{\rm{Tr}}[ {\bf{G}}_k^H( {{\bf{W}}_I} + {{\bf{W}}_E} + {\bf{V}}){{\bf{G}}_k}]$, ${\mu _k} \ge 0$ is the energy weight for the \textit{k}th ER, $0 < {\eta _k} \le 1$ is the energy conversion efficiency and we have \cite{IEEEhowto:15},\cite{IEEEhowto:33}
\begin{equation}\label{eq:F22}
\begin{array}{l}
{R_s} = \underbrace {\ln \left| {{\bf{I}} + {{\left[ {{\bf{I}} + {{\bf{H}}^H}\left( {{{\bf{W}}_E} + {\bf{V}}} \right){\bf{H}}} \right]}^{ - 1}}{{\bf{H}}^H}{{\bf{W}}_I}{\bf{H}}} \right|}_{{C_I}\left( {{{\bf{W}}_I}, {{\bf{W}}_E}, {\bf{V}}} \right)}\\
 ~~~~~~- \mathop {\max }\limits_{k \in {\rm K}} \underbrace {\ln \left| {{\bf{I}} + {{\left[ {{\bf{I}} + {\bf{G}}_k^H{\bf{V}}{{\bf{G}}_k}} \right]}^{ - 1}}{\bf{G}}_k^H{{\bf{W}}_I}{{\bf{G}}_k}} \right|}_{{C_{E,k}}\left( {{{\bf{W}}_I}, {{\bf{W}}_E}, {\bf{V}}} \right)}.
\end{array}
\end{equation}
The formulated WS-EHM problem (P3) is much more complex than the EHM problems mentioned in previous sections. In the following subsections, we will show how problem (P3) can be solved by our Tailor-made algorithm.

\subsection{Barrier Reformulation and GP Based Block GS Approach to Problem (P3)}

It can be seen that the secrecy rate constraint (\ref{eq:F21b}) is non-convex and it is difficult to deal with it straightly. Therefore, we need the aid of the following lemma.
\newtheorem{Lemma}{\textit{\underline{Lemma}}}[section]
\begin{Lemma}\label{lem:1}
Let N be any integer and matrix ${\bf{E}} \in {\mathbb{C}^{N \times N}}$ is arbitrary such that ${\bf{E}} \succ {\bf{0}}$. Consider the function $f\left( {\bf{T}} \right) = \ln \left| {\bf{T}} \right| - {\rm{Tr}}\left( {{\bf{TE}}} \right) + N$. Then,
\[\ln \left| {{{\bf{E}}^{ - 1}}} \right| = \mathop {\max }\limits_{{\bf{T}} \in {\mathbb{C}^{N \times N}}, {\bf{T}} \succ {\bf{0}}} f\left( {\bf{T}} \right),\]
with the optimal solution ${{\bf{T}}^*} = {{\bf{E}}^{ - 1}}$.
\end{Lemma}

Using Lemma \ref{lem:1}, by setting ${\bf{E}} = {\bf{I}} + {{\bf{H}}^H}\left( {{{\bf{W}}_E} + {\bf{V}}} \right){\bf{H}}$, ${C_I}\left( {{{\bf{W}}_I}, {{\bf{W}}_E}, {\bf{V}}} \right)$ can be rewritten as
\begin{equation}\label{eq:F23}
\begin{array}{l}
{C_I} = \ln \left| {{{\left[ {{\bf{I}} + {{\bf{H}}^H}\left( {{{\bf{W}}_E} + {\bf{V}}} \right){\bf{H}}} \right]}^{ - 1}}} \right|\\~~~~~~ + \ln \left| {{\bf{I}} + {{\bf{H}}^H}\left( {{{\bf{W}}_I} + {{\bf{W}}_E} + {\bf{V}}} \right){\bf{H}}} \right|\\
      {\kern 1pt}~~~= \mathop {\max }\limits_{{{\bf{T}}_0} \succeq {\bf{0}}} \ln \left| {{{\bf{T}}_0}} \right| - {\rm{Tr}}\left\{ {{{\bf{T}}_0}\left[ {{\bf{I}} + {{\bf{H}}^H}\left( {{{\bf{W}}_E} + {\bf{V}}} \right){\bf{H}}} \right]} \right\}\\ ~~~~~~+ {N_i}
            + \ln \left| {{\bf{I}} + {{\bf{H}}^H}\left( {{{\bf{W}}_I} + {{\bf{W}}_E} + {\bf{V}}} \right){\bf{H}}} \right|\\
  ~~~{\kern 1pt}   = \mathop {\max }\limits_{{{\bf{T}}_0} \succeq {\bf{0}}} {\theta _I}\left( {{{\bf{W}}_I}, {{\bf{W}}_E}, {\bf{V}}, {{\bf{T}}_0}} \right),
\end{array}
\end{equation}
where ${{\bf{T}}_0} \in {\mathbb{H}}_ + ^{{N_i}}$ and ${C_I}\left( {{{\bf{W}}_I}, {{\bf{W}}_E}, {\bf{V}}} \right)$ is simplified to ${C_I}$ for the convenience of description. Utilizing Lemma \ref{lem:1} again, set ${{\bf{E}}_k} = {\bf{I}} + {\bf{G}}_k^H\left( {{{\bf{W}}_I} + {{\bf{W}}_E} + {\bf{V}}} \right){{\bf{G}}_k}, \forall k \in {\rm K}$, we can rewrite ${C_{E,k}}\left( {{{\bf{W}}_I}, {{\bf{W}}_E}, {\bf{V}}} \right)$ as
\begin{equation}\label{eq:F24}
\begin{array}{l}
{C_{E,k}} = \ln |{[{\bf{I}} + {\bf{G}}_k^H{\bf{V}}{{\bf{G}}_k}]^{ - 1}}| + \ln |{\bf{I}} + {\bf{G}}_k^H({{\bf{W}}_I} + {\bf{V}}){{\bf{G}}_k}|\\
   ~~~~~~= \mathop {\min }\limits_{{{\bf{T}}_k} \succeq 0}  - \ln \left| {{\bf{I}} + {\bf{G}}_k^H{\bf{V}}{{\bf{G}}_k}} \right| - \ln \left| {{{\bf{T}}_k}} \right|\\
 ~~~~~~~~~+ {\rm{Tr}}\left\{ {{{\bf{T}}_k}\left[ {{\bf{I}} + {\bf{G}}_k^H\left( {{{\bf{W}}_I} + {\bf{V}}} \right){{\bf{G}}_k}} \right]} \right\} - {N_{e,k}}\\
   ~~~~~~= \mathop {\min }\limits_{{{\bf{T}}_k} \succeq 0} {\theta _{E,k}}\left( {{{\bf{W}}_I}, {{\bf{W}}_E}, {\bf{V}}, {{\bf{T}}_k}} \right), \forall k \in K,
\end{array}
\end{equation}
where ${{\bf{T}}_k} \in {\mathbb{H}}_ + ^{{N_e}}$ and ${C_{E,k}} = {C_{E,k}}\left( {{{\bf{W}}_I}, {{\bf{W}}_E}, {\bf{V}}} \right)$. By substituting (\ref{eq:F23}) and (\ref{eq:F24}) into (\ref{eq:F22}), we reformulate problem (P3) as
\begin{equation}\label{eq:F25}
\begin{array}{l}
\left( {\widetilde {\textsc{P3}}} \right) :    ~~    \mathop {\max }\limits_{\scriptstyle{{\bf{W}}_I} \succeq {\bf{0}}, {{\bf{W}}_E} \succeq {\bf{0}},\hfill\atop
\scriptstyle    ~ ~~~~  {\bf{V}} \succeq {\bf{0}}\hfill}  {{\rm{E}}_{{\rm{WS}}}}\left( {{{\bf{W}}_I}, {{\bf{W}}_E}, {\bf{V}}} \right)\\
          s.t. ~   \mathop {\max }\limits_{{{\bf{T}}_0} \succeq {\bf{0}}, {{\bf{T}}_k} \succeq {\bf{0}}} {\theta _I}\left( {{{\bf{W}}_I}, {{\bf{W}}_E}, {\bf{V}}, {{\bf{T}}_0}} \right)\\
 ~~~~~~~- {\theta _{E,k}}\left( {{{\bf{W}}_I}, {{\bf{W}}_E}, {\bf{V}}, {{\bf{T}}_k}} \right) \ge {R_0}\ln 2, \forall k \in {\rm K},\\
            ~~~~~~~~~~~~~~{\rm{Tr}}\left( {{{\bf{W}}_I} + {{\bf{W}}_E} + {\bf{V}}} \right) \le P.
\end{array}
\end{equation}

From (\ref{eq:F25}), we can see that a sequence of auxiliary variables $\left\{ {{{\bf{T}}_k}} \right\}_{k = 0}^K$ are introduced in reformulation $\left( {\widetilde {\textsc{P3}}} \right)$. Considering the existence of coupled variables and the feasible set is closed, nonempty and convex, we employ a block GS method [30] consists in solving two or three subproblems iteratively to settle problem $\left( {\widetilde {\textsc{P3}}} \right)$. Specifically, let $\left( {{\bf{W}}_I^n,{\kern 1pt} {\kern 1pt} {\bf{W}}_E^n,{{\bf{V}}^n},\left\{ {{\bf{T}}_k^n} \right\}_{k = 0}^K} \right)$ be the iterate at the nth iteration, each iteration of the block GS method consists of following subproblems:
\begin{equation}\label{eq:F26}
\left( {\widetilde {\textsc{P3}}-1} \right): {\bf{T}}_0^n = \arg \mathop {\max }\limits_{{{\bf{T}}_0} \succeq {\bf{0}}} {\theta _I}\left( {{\bf{W}}_I^{n - 1},{\bf{W}}_E^{n - 1},{{\bf{V}}^{n - 1}},{{\bf{T}}_0}} \right),
\end{equation}
\begin{equation}\label{eq:F27}
\begin{array}{l}
\left( {\widetilde {\textsc{P3}}-2} \right):~~~~ {\bf{T}}_k^n =\\
\arg \mathop {\max }\limits_{{{\bf{T}}_k} \ge {\bf{0}}} {\theta _{E,k}}\left( {{\bf{W}}_I^{n - 1},{\bf{W}}_E^{n - 1},{{\bf{V}}^{n - 1}},{{\bf{T}}_k}} \right),{\kern 1pt} {\kern 1pt} \forall k \in {\rm K},
\end{array}
\end{equation}
\begin{equation}\label{eq:F28}
\begin{array}{l}
\left( {\widetilde {\textsc{P3}}-3} \right):~~\left( {{\bf{W}}_I^n,{\kern 1pt} {\kern 1pt} {\bf{W}}_E^n,{{\bf{V}}^n}} \right) =\\
 \arg \mathop {\max }\limits_{\left( {{{\bf{W}}_I},{{\bf{W}}_E},{\bf{V}}} \right) \in {\bf{\Theta }}} {{\rm{E}}_{{\rm{WS}}}}\left( {{{\bf{W}}_I}, {{\bf{W}}_E}, {\bf{V}}} \right),
\end{array}
\end{equation}
where ${\bf{\Theta }} = {\rm{\{ }}\left( {{{\bf{W}}_I}, {{\bf{W}}_E}, {\bf{V}}} \right){\rm{|}}\;{\theta _I}\left( {{{\bf{W}}_I}, {{\bf{W}}_E}, {\bf{V}}, {\bf{T}}_{_0}^n} \right) - {\theta _{E,k}}\\\left( {{{\bf{W}}_I}, {{\bf{W}}_E}, {\bf{V}}, {\bf{T}}_k^n} \right) \ge {R_0}\ln 2, \forall k \in {\rm K},{\rm{Tr}}\left( {{\bf{W}}_I} + {{\bf{W}}_E} + {\bf{V}} \right)\\ \le P, {{\bf{W}}_I} \succeq {\bf{0}}, {{\bf{W}}_E} \succeq {\bf{0}}, {\bf{V}} \succeq {\bf{0}}{\rm{\} }}$, for $n = 1, 2,  \ldots $. In the light of Lemma \ref{lem:1}, given the value of $\left( {{{\bf{W}}_I},{{\bf{W}}_E},{\bf{V}}} \right)$ in the previous iteration, the solutions of subproblems $\left( {\widetilde {\textsc{P3}}-1} \right)$ and $\left( {\widetilde {\textsc{P3}}-2} \right)$  in the next iteration can be easily given by
\begin{equation}\label{eq:F29}
{\bf{T}}_0^n = {\left[ {{\bf{I}} + {{\bf{H}}^H}\left( {{\bf{W}}_E^{n - 1} + {{\bf{V}}^{n - 1}}} \right){\bf{H}}} \right]^{ - 1}},
\end{equation}
\begin{equation}\label{eq:F30}
{\bf{T}}_k^n = {\left[ {{\bf{I}} + {\bf{G}}_k^H\left( {{\bf{W}}_{_I}^{n - 1} + {{\bf{V}}^{n - 1}}} \right){{\bf{G}}_k}} \right]^{ - 1}}, \forall k \in {\rm K}.
\end{equation}
For subproblem $\left( {\widetilde {\textsc{P3}}-3} \right)$, since it is convex w.r.t. variables ${{\bf{W}}_I}$, ${{\bf{W}}_E}$ and ${\bf{V}}$ when ${\bf{T}}_0^n$ and $\left\{ {{\bf{T}}_k^n} \right\}_{k = 1}^K$ are fixed, the state of the art solver, such as the SeDuMi, can be used to solve this problem. Therefore, with such a conventional mind, we will solve a sequence of SDPs during the iteration. Here, instead, we put forward an alternative method in solving $\left( {\widetilde {\textsc{P3}}-3} \right)$ with much higher computational efficiency than the general method (cf. Table \ref{table_2}). We will first perform a logarithmic barrier reformulation of subproblem $\left( {\widetilde {\textsc{P3}}-3} \right)$, and then design a high-efficiency algorithm for the reformulated problem on the basis of GP method.

Two inequality constraints are included in $\left( {\widetilde {\textsc{P3}}-3} \right)$, to facilitate the subsequent process with GP method, we need to get rid of the secrecy rate constraint while no substantial change will be occurred to our problem. To this end, we resort to such an indicator function
\begin{equation}\label{eq:F31}
{I_{\rm{ + }}}\left( {x_k^n} \right) = \left\{ {{\kern 1pt} \begin{array}{*{20}{c}}
{{\kern 1pt} {\kern 1pt} {\kern 1pt} {\kern 1pt} 0,{\kern 1pt} {\kern 1pt} {\kern 1pt} {\kern 1pt} {\kern 1pt} {\kern 1pt} {\kern 1pt} {\kern 1pt} {\kern 1pt} {\kern 1pt} {\kern 1pt} {\kern 1pt} {\kern 1pt} {\kern 1pt} {\kern 1pt} {\kern 1pt} {\kern 1pt} {\kern 1pt} {\kern 1pt} {\kern 1pt} {\kern 1pt} {\kern 1pt} x_k^n \ge 0,{\kern 1pt} {\kern 1pt} }\\
{ - \infty ,{\kern 1pt} {\kern 1pt} {\kern 1pt} {\kern 1pt} {\kern 1pt} {\kern 1pt} {\kern 1pt} {\kern 1pt} {\kern 1pt} {\kern 1pt} {\kern 1pt} {\kern 1pt} {\kern 1pt} {\kern 1pt} x_k^n < 0,}
\end{array}} \right.
\end{equation}
where ${I_ + }: \mathbb{R}\to \mathbb{R}$ is the indicator function for the non-negative reals and $x_k^n = {x_k}\left( {{{\bf{W}}_I}, {{\bf{W}}_E}, {\bf{V}}, \left\{ {{\bf{T}}_{_k}^n} \right\}_{k = 0}^K} \right) = {\theta _I}\left( {{{\bf{W}}_I}, {{\bf{W}}_E}, {\bf{V}}, {\bf{T}}_{_0}^n} \right) - {\theta _{E,k}}\left( {{{\bf{W}}_I}, {{\bf{W}}_E}, {\bf{V}}, {\bf{T}}_k^n} \right) - {R_0}\ln 2$, $\forall k \in {\rm K}$. With this function, the secrecy rate constraint in subproblem $\left( {\widetilde {\textsc{P3}}-3} \right)$ can be removed to the objective and acts implicitly. When $x_k^n \ge 0$ holds during the iteration, the solution of the reformulation will be consistent with our original problem, if not, the objective value will be punished to negative infinity. However, it is obvious that the piecewise function in (\ref{eq:F31}) is non-differentiable. As a consequence, we introduce the logarithmic barrier function to approximate the indicator function\footnotetext[3]{The accuracy of the approximation is determined by the parameter $t$. As $t$ increase, $\widehat {{I_{\rm{ + }}}}$ will be closer to ${I_ + }$, i.e., $\widehat {{I_{\rm{ + }}}}\left( {x_k^n} \right)$ will be closer to zero when $x_k^n \ge 0$ is satisfied. The difference between both functions disappear once the value of $t$ tend to be infinite, i.e., $t \to \infty $.}, which can be depicted as:
\begin{equation}\label{eq:F32}
 \widehat {{I_{\rm{ + }}}}\left( {x_k^n} \right){\rm{ = }}\frac{1}{t}\ln \left( {x_k^n} \right),        {\rm{dom}}\widehat {{I_{\rm{ + }}}} = {\mathbb{R}_{ +  + }},
\end{equation}
where $t > 0$ is a parameter that sets the accuracy of the approximation. Note that $x_k^n$ is concave w.r.t. $\left( {{{\bf{W}}_I}, {{\bf{W}}_E}, {\bf{V}}} \right)$ and the logarithmic operation in (\ref{eq:F32}) is concave and non-decreasing, thus, according to the concept of convexity preserving operation \cite{IEEEhowto:32}, we conclude that the logarithmic barrier function $\widehat {{I_{\rm{ + }}}}\left( {x_k^n} \right)$ is concave w.r.t. $\left( {{{\bf{W}}_I}, {{\bf{W}}_E}, {\bf{V}}} \right)$.

Making (\ref{eq:F32}) in the objective of subproblem $\left( {\widetilde {\textsc{P3}}-3} \right)$, we get the reformulation
\begin{equation}\label{eq:F33}
\begin{array}{l}
\left( {\widetilde {\textsc{P3}}-3\textsc{A}} \right):    {\kern 1pt} \left( {{\bf{W}}_I^n,{\kern 1pt} {\kern 1pt} {\bf{W}}_E^n,{{\bf{V}}^n}} \right) = \\
\arg \mathop {\max }\limits_{\left( {{{\bf{W}}_I},{{\bf{W}}_E},{\bf{V}}} \right) \in {\bf{\Xi }}} {{\rm{E}}_{{\rm{WS}}}}\left( {{{\bf{W}}_I}, {{\bf{W}}_E}, {\bf{V}}} \right) + \frac{1}{t}\sum\limits_{k = 1}^K {\ln \left( {x_k^n} \right)}
\end{array}
\end{equation}
where ${\bf{\Xi }} = {\rm{\{ }}\left( {{{\bf{W}}_I},{{\bf{W}}_E},{\bf{V}}} \right){\rm{|\;Tr}}\left( {{{\bf{W}}_I} + {{\bf{W}}_E} + {\bf{V}}} \right) \le P, {{\bf{W}}_I}\\ \succeq {\bf{0}}, {{\bf{W}}_E} \succeq {\bf{0}}, {\bf{V}} \succeq {\bf{0}}{\rm{\} }}$ and the Cumulative-Sum operation on the logarithmic barrier function is due to the existence of \textit{K} secrecy rate constraints.

To provide a solver to problem $\left( {\widetilde {\textsc{P3}}-3\textsc{A}} \right)$, we turn to the GP method which possesses low computational cost by exploiting the problem structure. Let $g\left( {{{\bf{W}}_I},{{\bf{W}}_E},{\bf{V}}} \right) = {{\rm{E}}_{{\rm{WS}}}}\left( {{{\bf{W}}_I}, {{\bf{W}}_E}, {\bf{V}}} \right) + \frac{1}{t}\sum\limits_{k = 1}^K {\ln \left( {x_k^n} \right)} $. The GP method applied to problem $\left( {\widetilde {\textsc{P3}}-3\textsc{A}} \right)$ is defined as follows \cite{IEEEhowto:31}:
\begin{equation}\label{eq:F34}
\begin{array}{l}
\left( {\widehat {\bf{W}}_I^{n, m + 1}, \widehat {\bf{W}}_{_E}^{n, m + 1}, {{\widehat {\bf{V}}}^{n,m + 1}}} \right)\\ = {\rm{Pro}}{{\rm{j}}_{\bf{\Xi }}}[\left( {{\bf{W}}_I^{n, m}, {\bf{W}}_E^{n, m}, {{\bf{V}}^{n, m}}} \right)\\
~~~~~~~~~~~ + {q_1}\left( {{\nabla _{{{\bf{W}}_I}}}{g^{n, m}}, {\nabla _{{{\bf{W}}_E}}}{g^{n, m}}, {\nabla _{\bf{V}}}{g^{n, m}}} \right),
\end{array}
\end{equation}
\begin{equation}\label{eq:F35}
\begin{array}{l}
\left( {{\bf{W}}_I^{n, m + 1},{\bf{W}}_E^{n, m + 1},{{\bf{V}}^{n, m + 1}}} \right) = \left( {{\bf{W}}_I^{n, m},{\bf{W}}_E^{n, m},{{\bf{V}}^{n, m}}} \right)+\\
{q_2}\left( {( {\widehat {\bf{W}}_I^{n, m + 1},\widehat {\bf{W}}_E^{n, m + 1},{{\widehat {\bf{V}}}^{n, m + 1}}} ) - ( {{\bf{W}}_I^{n, m},{\bf{W}}_E^{n, m},{{\bf{V}}^{n, m}}})} \right)
\end{array}
\end{equation}
where $\left( {\widehat {\bf{W}}_I^{n, m + 1},\widehat {\bf{W}}_{_E}^{n, m + 1},{{\widehat {\bf{V}}}^{n,m + 1}}} \right)$ is the introduced auxiliary variable set during the iteration \textit{m+1} of GP method; ${\rm{Pro}}{{\rm{j}}_{\bf{\Xi }}}$ denotes the projection on feasible set $\bm{\Xi} $ and its analytical expression is depicted as:
\begin{equation}\label{eq:F36}
\begin{array}{l}
{\rm{Pro}}{{\rm{j}}_{\bf{\Xi }}}\left( {{{{\bf{\bar W}}}_I}, {{{\bf{\bar W}}}_E}, {\bf{\bar V}}} \right) = \arg  \mathop {\min }\limits_{\left( {{{\widehat {\bf{W}}}_I},{{\widehat {\bf{W}}}_E},\widehat {\bf{V}}} \right) \in {\bf{\Xi }}} \left\| {{{\widehat {\bf{W}}}_I} - {{{\bf{\bar W}}}_I}} \right\|_F^2\\
 ~~~~~~~~~~~~~~~~~~~~~~~~~~~+ \left\| {{{\widehat {\bf{W}}}_E} - {{{\bf{\bar W}}}_E}} \right\|_F^2 + \left\| {\widehat {\bf{V}} - {\bf{\bar V}}} \right\|_F^2;
\end{array}
\end{equation}
${q_1} > 0$ and $0 < {q_2} \le 1$ are the step sizes, and one can refer to the Armijo rule \cite{IEEEhowto:31} for the details about the step size selection; $g$, a shorter version, denotes $g\left( {{{\bf{W}}_I},{{\bf{W}}_E},{\bf{V}}} \right)$ and $\left( {{\nabla _{{{\bf{W}}_I}}}{g^{n, m}}, {\nabla _{{{\bf{W}}_E}}}{g^{n, m}}, {\nabla _{\bf{V}}}{g^{n, m}}} \right)$ denotes the gradient of $g\left( {{{\bf{W}}_I},{{\bf{W}}_E},{\bf{V}}} \right)$ w.r.t. $\left( {{\bf{W}}_I^{n, m}, {\bf{W}}_E^{n, m}, {{\bf{V}}^{n, m}}} \right)$, which is given particularly in (\ref{eq:F37}) on the top of next page, where ${{\bf{\Omega }}^{n,{\kern 1pt} m}} = {\bf{H}}{\left[ {{\bf{I}} + {{\bf{H}}^H}\left( {{\bf{W}}_I^{n,{\kern 1pt} m} + {\bf{W}}_E^{n,{\kern 1pt} m} + {{\bf{V}}^{n,{\kern 1pt} m}}} \right){\bf{H}}} \right]^{ - 1}}{{\bf{H}}^H}$ and ${{\bf{\Psi }}^{n,{\kern 1pt} m}} = {{\bf{G}}_k}{\left[ {{\bf{I}} + {\bf{G}}_k^H{{\bf{V}}^{n,{\kern 1pt} m}}{{\bf{G}}_k}} \right]^{ - 1}}{\bf{G}}_k^H$, $\forall k \in {\rm K}$.
\begin{figure*}[!t]
\normalsize
\begin{equation}\label{eq:F37}
\left\{ \begin{array}{l}
{\nabla _{{{\bf{W}}_I}}}{g^{n, m}} = \sum\limits_{k = 1}^K {\sigma _{e,k}^2} {\mu _k}{\eta _k}{{\bf{G}}_k}{\bf{G}}_k^H + \frac{1}{t}\sum\limits_{k = 1}^K {\frac{1}{{x_k^n}}\left( {{{\bf{\Omega }}^{n, m}} - {{\bf{G}}_k}{\bf{T}}_k^n{\bf{G}}_k^H} \right),} \\
{\nabla _{{{\bf{W}}_E}}}{g^{n, m}} = \sum\limits_{k = 1}^K {\sigma _{e,k}^2} {\mu _k}{\eta _k}{{\bf{G}}_k}{\bf{G}}_k^H + \frac{1}{t}\sum\limits_{k = 1}^K {\frac{1}{{x_k^n}}\left( { - {\bf{HT}}_0^n{{\bf{H}}^H} + {{\bf{\Omega }}^{n, m}}} \right),} \\
{\nabla _{\bf{V}}}{g^{n, m}} = \sum\limits_{k = 1}^K {\sigma _{e,k}^2} {\mu _k}{\eta _k}{{\bf{G}}_k}{\bf{G}}_k^H + \frac{1}{t}\sum\limits_{k = 1}^K {\frac{1}{{x_k^n}}\left( { - {\bf{HT}}_0^n{{\bf{H}}^H} + {{\bf{\Omega }}^{n, m}} + {{\bf{\Psi }}^{n, m}} - {{\bf{G}}_k}{\bf{T}}_k^n{\bf{G}}_k^H} \right),}
\end{array} \right.
\end{equation}
\hrulefill
\vspace*{4pt}
\end{figure*}

The core of the GP method lies in the projection process in (\ref{eq:F36}), by taking full advantage of the structure of feasible region ${\bf{\Xi }}$, we can implement the projection effectively.
\newtheorem{prop}{\textit{\underline{Proposition}}}[section]
\begin{Proposition}\label{pro:3}
 Similar to the skills, SVD and EVD, applied in dealing with subproblem $\left( {\widetilde {\textsc{P1}} - 2} \right)$, carry out EVD of each known quantity in (\ref{eq:F36}), i.e., ${{\bf{\bar W}}_I} = {{\bf{U}}_{{{\bf{W}}_I}}}{{\bf{\bar \Lambda }}_{{{\bf{W}}_I}}}{\bf{U}}_{{{\bf{W}}_I}}^H$, ${{\bf{\bar W}}_E} = {{\bf{U}}_{{{\bf{W}}_E}}}{{\bf{\bar \Lambda }}_{{{\bf{W}}_E}}}{\bf{U}}_{{{\bf{W}}_E}}^H$ and ${\bf{\bar V}} = {{\bf{U}}_{\bf{V}}}{{\bf{\bar \Lambda }}_{\bf{V}}}{\bf{U}}_{\bf{V}}^H$. Then, we have
\[\begin{array}{l}
{\rm{Pro}}{{\rm{j}}_{\bf{\Xi }}}\left( {{{{\bf{\bar W}}}_I}, {{{\bf{\bar W}}}_E}, {\bf{\bar V}}} \right) = \left( {{{\widehat {\bf{W}}}_I}, {{\widehat {\bf{W}}}_E}, \widehat {\bf{V}}} \right)\\
~~~~= \left( {{{\bf{U}}_{{{\bf{W}}_I}}}{{\widehat {\bf{\Lambda }}}_{{{\bf{W}}_I}}}{\bf{U}}_{{{\bf{W}}_I}}^H, {{\bf{U}}_{{{\bf{W}}_E}}}{{\widehat {\bf{\Lambda }}}_{{{\bf{W}}_E}}}{\bf{U}}_{{{\bf{W}}_E}}^H, {{\bf{U}}_{\bf{V}}}{{\widehat {\bf{\Lambda }}}_{\bf{V}}}{\bf{U}}_{\bf{V}}^H} \right).
\end{array}\]
Motivated by this, we now turn our attention to the calculation of $\left( {{{\widehat {\bf{\Lambda }}}_{{{\bf{W}}_I}}}, {{\widehat {\bf{\Lambda }}}_{{{\bf{W}}_E}}}, {{\widehat {\bf{\Lambda }}}_{\bf{V}}}} \right)$ instead of $\left( {{{\widehat {\bf{W}}}_I}, {{\widehat {\bf{W}}}_E}, \widehat {\bf{V}}} \right)$.
\begin{equation}\label{eq:F38}
\left( {\widehat {\bf{\Lambda }}_{_{{{\bf{W}}_I}}}^*, \widehat {\bf{\Lambda }}_{_{{{\bf{W}}_E}}}^{\rm{*}}, \widehat {\bf{\Lambda }}_{_{\bf{V}}}^{\rm{*}}} \right){\rm{ = }}\arg \mathop {\min }\limits_{\left( {{{\widehat {\bf{\Lambda }}}_{{{\bf{W}}_I}}}, {{\widehat {\bf{\Lambda }}}_{{{\bf{W}}_E}}}, {{\widehat {\bf{\Lambda }}}_{\bf{V}}}} \right) \in {\bf{\Xi }}} \left\| {\widehat {\bf{\Lambda }} - {\bf{\bar \Lambda }}} \right\|_F^2
\end{equation}
\end{Proposition}
where $\widehat {\bf{\Lambda }}{\rm{ = }}\left( {{\rm{diag}}( {{{\widehat {\bf{\Lambda }}}_{{{\bf{W}}_I}}}} ),{\rm{diag}}( {{{\widehat {\bf{\Lambda }}}_{{{\bf{W}}_E}}}} ),{\rm{diag}}( {{{\widehat {\bf{\Lambda }}}_{\bf{V}}}} )} \right)$ and ${\bf{\bar \Lambda }}{\rm{ = }}\\\left( {{\rm{diag}}\left( {{{{\bf{\bar \Lambda }}}_{{{\bf{W}}_I}}}} \right),{\rm{diag}}\left( {{{{\bf{\bar \Lambda }}}_{{{\bf{W}}_E}}}} \right),{\rm{diag}}\left( {{{{\bf{\bar \Lambda }}}_{\bf{V}}}} \right)} \right)$.

\textit{Proof:} First of all, by writing ${{\bf{\bar W}}_I}$, ${{\bf{\bar W}}_E}$ and ${\bf{\bar V}}$ in their EVD forms, we obtain the equivalent transformation of (\ref{eq:F36}) which is shown in (\ref{eq:F39}), where ${\widehat {\bf{\Lambda }}_{{{\bf{W}}_I}}}$, ${\widehat {\bf{\Lambda }}_{{{\bf{W}}_E}}}$ and ${\widehat {\bf{\Lambda }}_{\bf{V}}}$ are defined as ${\bf{U}}_{{{\bf{W}}_I}}^H{\widehat {\bf{W}}_I}{{\bf{U}}_{{{\bf{W}}_I}}}$, ${\bf{U}}_{{{\bf{W}}_E}}^H{\widehat {\bf{W}}_E}{{\bf{U}}_{{{\bf{W}}_E}}}$ and ${\bf{U}}_{\bf{V}}^H\widehat {\bf{V}}{{\bf{U}}_{\bf{V}}}$, resp.. Our next endeavor is to certify that the optimal solution $\left( {\widehat {\bf{\Lambda }}_{_{{{\bf{W}}_I}}}^*, \widehat {\bf{\Lambda }}_{_{{{\bf{W}}_E}}}^*, \widehat {\bf{\Lambda }}_{_{\bf{V}}}^*} \right)$ of (\ref{eq:F38}) must be diagonal. Assuming an optimal solution $\left( {{{\widehat {\bf{\Lambda }}}^\prime }_{{{\bf{W}}_I}}, {{\widehat {\bf{\Lambda }}}^\prime }_{{{\bf{W}}_E}}, {{\widehat {\bf{\Lambda }}}^\prime }_{\bf{V}}} \right)$ which is not diagonal and the resulting objective value is $y'$. According to (\ref{eq:F39}), it is easy to learn that there must have a diagonal counterpart which achieves a smaller objective than $y'$ in the feasible zone ${\bf{\Xi }}$. Noticing such a contradiction, we conclude that (\ref{eq:F36}) is equivalent to its simplified version (\ref{eq:F38}).
\begin{figure*}[!t]
\normalsize
\begin{equation}\label{eq:F39}
\begin{array}{l}
\mathop {\min }\limits_{\left( {{{\widehat {\bf{W}}}_I},{{\widehat {\bf{W}}}_E},\widehat {\bf{V}}} \right) \in {\bf{\Xi }}} \left\| {{{\widehat {\bf{W}}}_I} - {{\bf{U}}_{{{\bf{W}}_I}}}{{{\bf{\bar \Lambda }}}_{{{\bf{W}}_I}}}{\bf{U}}_{{{\bf{W}}_I}}^H} \right\|_F^2 + \left\| {{{\widehat {\bf{W}}}_E} - {{\bf{U}}_{{{\bf{W}}_E}}}{{{\bf{\bar \Lambda }}}_{{{\bf{W}}_E}}}{\bf{U}}_{{{\bf{W}}_E}}^H} \right\|_F^2 + \left\| {\widehat {\bf{V}} - {{\bf{U}}_{\bf{V}}}{{{\bf{\bar \Lambda }}}_{\bf{V}}}{\bf{U}}_{\bf{V}}^H} \right\|_F^2\\
 \Leftrightarrow \mathop {\min }\limits_{\left( {{{\widehat {\bf{W}}}_I},{{\widehat {\bf{W}}}_E},\widehat {\bf{V}}} \right) \in {\bf{\Xi }}} \left\| {{\bf{U}}_{{{\bf{W}}_I}}^H{{\widehat {\bf{W}}}_I}{{\bf{U}}_{{{\bf{W}}_I}}} - {{{\bf{\bar \Lambda }}}_{{{\bf{W}}_I}}}} \right\|_F^2 + \left\| {{\bf{U}}_{{{\bf{W}}_E}}^H{{\widehat {\bf{W}}}_E}{{\bf{U}}_{{{\bf{W}}_E}}} - {{{\bf{\bar \Lambda }}}_{{{\bf{W}}_E}}}} \right\|_F^2 + \left\| {{\bf{U}}_{\bf{V}}^H\widehat {\bf{V}}{{\bf{U}}_{\bf{V}}} - {{{\bf{\bar \Lambda }}}_{\bf{V}}}} \right\|_F^2\\
 \Leftrightarrow \mathop {\min }\limits_{\left( {{{\widehat {\bf{\Lambda }}}_{{{\bf{W}}_I}}}, {{\widehat {\bf{\Lambda }}}_{{{\bf{W}}_E}}}, {{\widehat {\bf{\Lambda }}}_{\bf{V}}}} \right) \in {\bf{\Xi }}} \left\| {{{\widehat {\bf{\Lambda }}}_{{{\bf{W}}_I}}} - {{{\bf{\bar \Lambda }}}_{{{\bf{W}}_I}}}} \right\|_F^2 + \left\| {{{\widehat {\bf{\Lambda }}}_{{{\bf{W}}_E}}} - {{{\bf{\bar \Lambda }}}_{{{\bf{W}}_E}}}} \right\|_F^2 + \left\| {{{\widehat {\bf{\Lambda }}}_{\bf{V}}} - {{{\bf{\bar \Lambda }}}_{\bf{V}}}} \right\|_F^2,
\end{array}
\end{equation}
\hrulefill
\vspace*{4pt}
\end{figure*}

In accordance with the argument in \cite{IEEEhowto:37}, the optimal solution of (\ref{eq:F38}) can be given by
\begin{equation}\label{eq:F40}
\begin{array}{l}
\widehat {\bf{\Lambda }}_{_{{{\bf{W}}_I}}}^* = {\left( {{{{\bf{\bar \Lambda }}}_{{{\bf{W}}_I}}} - {\rho ^*}{\bf{I}}} \right)^ + }, \widehat {\bf{\Lambda }}_{_{{{\bf{W}}_E}}}^* = {\left( {{{{\bf{\bar \Lambda }}}_{{{\bf{W}}_E}}} - {\rho ^*}{\bf{I}}} \right)^ + }, \\
~~~~~~~~~~~~~~~~~\widehat {\bf{\Lambda }}_{_{\bf{V}}}^* = {\left( {{{{\bf{\bar \Lambda }}}_{\bf{V}}} - {\rho ^*}{\bf{I}}} \right)^ + },
\end{array}
\end{equation}
where ${\rho ^*}$ can be calculated efficiently through the Algorithm \ref{alg:Algorithm 1} in \cite{IEEEhowto:37} and ${\left( x \right)^ + } \buildrel \Delta \over = \max \left( {0,{\kern 1pt} {\kern 1pt} {\kern 1pt} x} \right)$.

So far, we accomplish the derivation of our custom-made barrier-GP method for subproblem $\left( {\widetilde {\textsc{P3}} - 3} \right)$. The details of barrier-GP method together with the whole block GS process are summarized in Algorithm \ref{alg:Algorithm 3}.
\begin{algorithm}[!t]    
\caption{The barrier-GP based block GS algorithm for WS-EHM (P3).}   
\begin{algorithmic}[1] \label{alg:Algorithm 3}     
\STATE \textbf{Initialize:} $n=0$, ${q_1} > 0$, ${\mu _t} > 1$, ${\xi _1} > 0, \;{\xi _2} > 0,\; {\xi _3} > 0$, ${\bf{W}}_I^0 \succeq {\bf{0}},\;{\bf{W}}_E^0 \succeq {\bf{0}},\;{{\bf{V}}^0} \succeq {\bf{0}}$;\\
\STATE\textbf{Repeat}
  \STATE ~~$n=n+1$, $t = 1 $;
  \STATE ~~Calculate ${\bf{T}}_0^n$ and $\left\{ {{\bf{T}}_k^n} \right\}_{k = 1}^K$ according to (\ref{eq:F26}) and (\ref{eq:F27}), resp.;
  \STATE ~~Set $\left( {{\bf{W}}_I^{n,{\kern 1pt} 0},{\kern 1pt} {\kern 1pt} {\kern 1pt} {\bf{W}}_E^{n,{\kern 1pt} 0},{{\bf{V}}^{n,{\kern 1pt} 0}}} \right) = \left( {{\bf{W}}_I^{n - 1},{\bf{W}}_E^{n - 1},{{\bf{V}}^{n - 1}}} \right)$;
  \STATE ~~\textbf{Repeat}
  \STATE ~~~~$m=0$;
  \STATE ~~~~\textbf{Repeat}
  \STATE ~~~~~~Compute $\left( {{\nabla _{{{\bf{W}}_I}}}{g^{n, m}}, {\nabla _{{{\bf{W}}_E}}}{g^{n, m}}, {\nabla _{\bf{V}}}{g^{n, m}}} \right)$ in line with (\ref{eq:F37});
  \STATE ~~~~~~Compute $\left( {\widehat {\bf{W}}_I^{n, m + 1},\widehat {\bf{W}}_{_E}^{n, m + 1},{{\widehat {\bf{V}}}^{n,m + 1}}} \right)$ according to Proposition \ref{pro:3} and (\ref{eq:F40});
  \STATE ~~~~~~Compute $\left( {{\bf{W}}_I^{n, m}, {\bf{W}}_E^{n, m}, {{\bf{V}}^{n, m}}} \right)$ according to (\ref{eq:F35}), where the value of ${q_2}$ is determined by the Armijo rule in \cite{IEEEhowto:31};
  \STATE ~~~~~~$m=m+1$;
  \STATE ~~~~\textbf{Until} $\left| {{{\left( {{g^{n,{\kern 1pt} m}} - {g^{n,{\kern 1pt} m - 1}}} \right)} \mathord{\left/ {\vphantom {{\left( {{g^{n,{\kern 1pt} m}} - {g^{n,{\kern 1pt} m - 1}}} \right)} {{g^{n,{\kern 1pt} m - 1}}}}} \right. \kern-\nulldelimiterspace} {{g^{n,{\kern 1pt} m - 1}}}}} \right| < {\xi _1}$
  \STATE ~~~~Update $t = {\mu _t}{\kern 1pt} t$;
  \STATE ~~\textbf{Until} ${\raise0.7ex\hbox{$K$} \!\mathord{\left/{\vphantom {K t}}\right.\kern-\nulldelimiterspace}\!\lower0.7ex\hbox{$t$}} < {\xi _2}$
  \STATE ~~Update ${\bf{W}}_I^n = {\bf{W}}_I^{n,{\kern 1pt} m}$, ${\bf{W}}_E^n = {\bf{W}}_E^{n,{\kern 1pt} m}$, ${{\bf{V}}^n} = {{\bf{V}}^{n,{\kern 1pt} m}}$;
  \STATE \textbf{Until} $\left| {{\rm{E}}_{_{{\rm{WS}}}}^n - {\rm{E}}_{_{{\rm{WS}}}}^{n - 1}} \right| < {\xi _3}$
\STATE ${\bf{W}}_{^I}^* = {\bf{W}}_{^I}^n$, ${\bf{W}}_E^* = {\bf{W}}_{^E}^n$, ${{\bf{V}}^*} = {{\bf{V}}^n}$;
\STATE \textbf{Output:} ${\bf{W}}_{^I}^*$, ${\bf{W}}_{^E}^*$ and ${{\bf{V}}^*}$.
\end{algorithmic}
\end{algorithm}

Like the measures that we ever took in Algorithm \ref{alg:Algorithm 1}, here, the feasibility assurance procedure of the initial values ${\bf{W}}_I^0$, ${\bf{W}}_E^0$ and ${{\bf{V}}^0}$ is also necessary during each channel realization. It should be noted that in Algorithm \ref{alg:Algorithm 3}, the parameter $t$ is updated (increased) by a factor ${\mu _t} > 1$, under the guidance of central path which converges to the optimal point as $t \to \infty $ \cite{IEEEhowto:32}. Since the value of $t$ can be updated at each iteration and tends to be infinite (${\xi _2}$ is infinitesimal), the reformulation $\left( {\widetilde {\textsc{P3}} - 3{\rm{A}}} \right)$ can be deemed to be equivalent to $\left( {\widetilde {\textsc{P3}} - 3}\right)$. Therefore, the resulting solution will be optimal to problem $\left( {\widetilde {\textsc{P3}}}\right)$ as well as problem (P3).

\newtheorem{propo}{\textit{\underline{Proposition}}}[section]
\begin{Proposition}\label{pro:4}
The monotonically non-decreasing objective values are produced from our proposed barrier-GP based block GS algorithm. Moreover, every limit point $\left( {{\bf{W}}_{^I}^*,{\bf{W}}_{^E}^*,{{\bf{V}}^*}} \right)$ of the iterations generated by the block GS process is a KKT point of problem (P3).
\end{Proposition}

\textit{Proof:} Please refer to Appendix C.

At this point, we finished the study of AN-aided WS-EHM problem for the case that the IR is not able to eliminate the interference from the energy signals. Since the whole process presented above is still available for the problem dropping ${{\bf{W}}_E}$ from (\ref{eq:F22}), the research of the case in which the IR is able to remove the interference from the energy signals is omitted here.

\section{Numerical Results}

In this section, we provide numerical results to validate the performance of our proposed algorithms for EHM and WS-EHM. In each simulation trial, the distance-dependent path loss model is given by
\[\mathord{\buildrel{\lower3pt\hbox{$\scriptscriptstyle\smile$}}
\over L} {\rm{ = }}{{\rm{A}}_0}{\left( {\frac{d}{{{d_0}}}} \right)^{ - \gamma }},{\kern 1pt} {\kern 1pt} d > {d_0},\]
where ${{\rm{A}}_0}{\rm{ = }}1$, $d$ denotes the distance between the transmitter and the associated receiver (IR or ER), ${d_0}$ is a reference distance set to be 1m, and $\gamma $ is the path loss exponent set at 3.

\subsection{One ER EHM}

\begin{figure}[!t]
\centering
\includegraphics[width=3.2in]{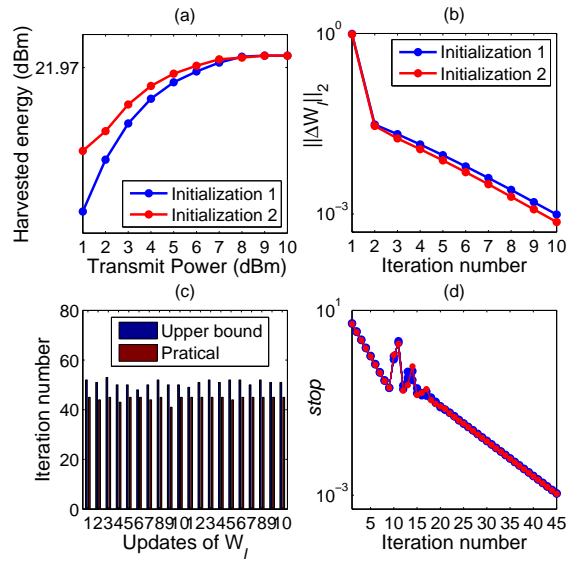}
\caption{Convergence analysis of Algorithm \ref{alg:Algorithm 1} for $P = 10 {\rm{dB}}$ and $\alpha  = 0.5$.}
\label{fig:simulation1}
\end{figure}

In this subsection, we consider the MIMO one ER scenario. The number of the antennas at the transmitter, IR and ER are ${N_t} = 5$, ${N_i} = 3$ and ${N_e} = 3$, resp.. The channels ${\bf{\mathord{\buildrel{\lower3pt\hbox{$\scriptscriptstyle\smile$}}\over H} }}$ and ${\bf{\mathord{\buildrel{\lower3pt\hbox{$\scriptscriptstyle\smile$}}\over G} }}$ are assumed to be quasi-static flat Rayleigh fading, and each element of them follows an independent complex Gaussian distribution with zero mean and a covariance specified by $\mathord{\buildrel{\lower3pt\hbox{$\scriptscriptstyle\smile$}}\over L} $, where the transmitter is assumed to have the distances ${d_i} = 9{\rm{m}}$ and ${d_e} = 7{\rm{m}}$ away from IR and ER, resp... Also, we set other parameters as: $\sigma _i^2 = \sigma _e^2 =  - 5 {\rm{dBm}}$, $\eta  = 0.8$, ${C_0} = 3 {\rm{bit/s/Hz}}$, $\lambda  = \mu  = 15$, ${r^2} = 10$, ${\varepsilon _1} = {\varepsilon _2} = \zeta  = {10^{ - 3}}$ and ${{\bf{W}}_{I0}} = {\bf{0}}$. Fig. \ref{fig:simulation1} shows the convergence performance of Algorithm \ref{alg:Algorithm 1} when fixing $P = 10 {\rm{dB}}$ and $\alpha  = 0.5$. Two different initial feasible points after the feasibility check process of ${{\bf{W}}_{I0}}$ are selected as the examples. The results show that the proposed EHM algorithm converges to the same harvested energy with different initializations (cf. Fig. \ref{fig:simulation1}(a)) and the initializations converge to optimal gradually after each iteration (cf. Fig. \ref{fig:simulation1}(b)). Fig. \ref{fig:simulation1}(c) and Fig. \ref{fig:simulation1}(d) show the convergence of the ellipsoid method adopted in updating the Lagrangian dual variables $\lambda $ and $\mu $. During each update of variable ${{\bf{W}}_I}$, the ellipsoid method is introduced to obtain the relevant optimal dual variables ${\lambda ^{\rm{*}}}$ and ${\mu ^{\rm{*}}}$. It is observed that the actual iteration numbers which fulfills the stop criterion is limited and less than the upper bound value (i.e., $8\ln \left( {\frac{{r{l_s}}}{\varepsilon }} \right)$).

\begin{figure}[!t]
\centering
\includegraphics[width=3.2in]{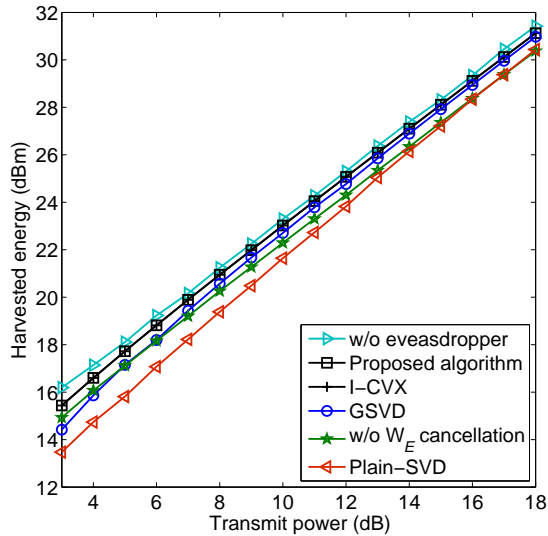}
\caption{Harvested energy versus the transmit power.}
\label{fig:simulation2}
\end{figure}

\begin{figure}[!t]
\centering
\includegraphics[width=3.2in]{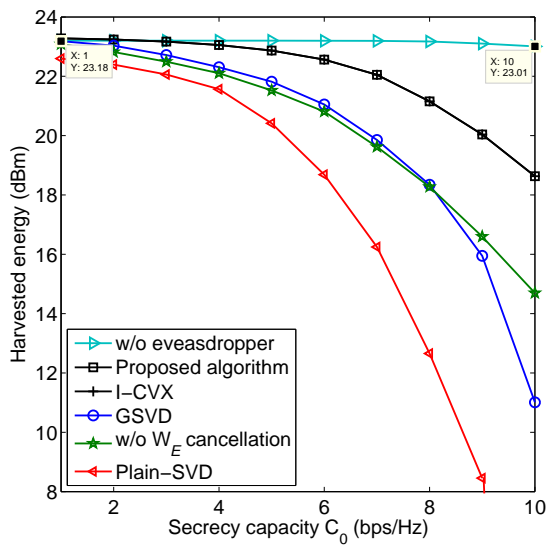}
\caption{Harvested energy versus the secrecy capacity.}
\label{fig:simulation3}
\end{figure}

Then, we show the energy harvesting performance of our proposed algorithm and compare it with some existing methods and scenarios, namely, the iterative CVX (I-CVX) method, the GSVD method \cite{IEEEhowto:9}, the Plain-SVD method \cite{IEEEhowto:16}, the ER not serving as an eavesdropper (w/o eavesdropper) and the IR not possessing the ability to perform interference cancellation (w/o ${{\bf{W}}_E}$ cancellation). Fig. \ref{fig:simulation2} shows the average harvested energy of various methods and different scenario setups w.r.t. the transmitter power. It can be seen that our proposed algorithm performs identical to the I-CVX method and better than the other methods in terms of energy harvesting. Regarding the w/o eavesdropper case, since there is no intercept, less power is required to fulfill the rate constraint, which, thereby, resulting in more power transfer to ER. On the contrary, when the energy signals act as the disturbance to IR (i.e., the w/o ${{\bf{W}}_E}$ cancellation case), more power is allocated to fulfill the secrecy capacity constraint, which, of course, leads to a worse energy harvesting performance for ER. Fig. \ref{fig:simulation3} shows the harvested energy versus the secrecy capacity constraint with fixed $P = 10{\rm{dB}}$. Again, our proposed algorithm yields the uniform performance with the I-CVX method and superior to other methods. In the w/o eavesdropper scenario, the harvested energy decreased gently with the increasing secrecy capacity requirement. The reason is that the lower bound of the desired rate of IR can be achieved easily so that a large proportion of transmitter power is assigned to transport energy for ER. In this figure, a trade-off design between the harvested energy at ER and secrecy capacity at IR is also provided for each method or scenario. Table \ref{table_1} shows the average running time of our proposed and I-CVX methods. It reveals intuitively that our proposed algorithm owns a much lower computational complexity than I-CVX.

\begin{table}[!t]

\caption{Average running time (in secs.) versus the transmit power}
\label{table_1}
\centering
\begin{tabular}{c|c|c|c|c|c|c}
\hline
\multirow{2}{*}{Method} & \multicolumn{6}{c}{Transmit Power (dB)}\\
\cline{2-7}
&3&6&9&12&15&18\\
\hline
Proposed&0.0875&0.1067&0.1339&0.1526&0.1877&0.2136\\
I-CVX&2.0628&2.4692&2.8700&3.3599&3.9975&5.8160\\
\hline
\end{tabular}
\end{table}

\subsection{Multiple ERs WS-EHM}

\begin{figure}[!t]
\centering
\includegraphics[width=3.2in]{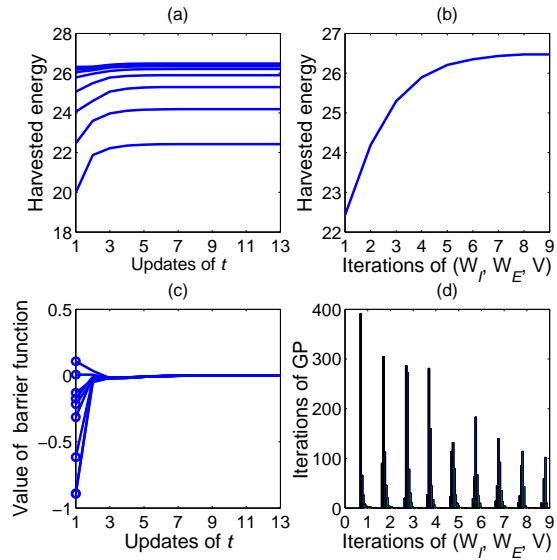}
\caption{Convergence analysis of Algorithm \ref{alg:Algorithm 3}.}
\label{fig:simulation4}
\end{figure}

In this subsection, we consider the MIMO multiple ERs scenario. The channels ${\bf{\mathord{\buildrel{\lower3pt\hbox{$\scriptscriptstyle\smile$}}
\over H} }}$ and ${{\bf{\mathord{\buildrel{\lower3pt\hbox{$\scriptscriptstyle\smile$}}
\over G} }}_k}, k \in {\rm K}$, are the same as in previous subsection while ${d_i} = 9m$ and ${d_{e, k}} = 7m, \forall k$. The other parameters are set as: ${N_t} = 5$, ${N_i} = 3$, ${N_{e, k}} = 3, \forall k$, $K = 3$, $\sigma _i^2 = \sigma _{e, k}^2 =  - 5 {\rm{dBm}}$, ${\mu _k} = 1, \forall k$, ${\eta _k} = 0.8, \forall k$, ${\mu _t} = 3$, ${q_1} = 0.1P$, ${\xi _1} = {\xi _3} = {10^{ - 3}}$, ${\xi _2} = {10^{ - 6}}$ and ${\bf{W}}_I^0 = {\bf{W}}_E^0 = {{\bf{V}}^0} = \left( {{P \mathord{\left/
 {\vphantom {P {3{N_t}}}} \right.
 \kern-\nulldelimiterspace} {3{N_t}}}} \right){\bf{I}}$. Fig. \ref{fig:simulation4} shows the convergence performance of Algorithm \ref{alg:Algorithm 3} when fixing $P = 10{\rm{dB}}$ and ${R_0} = 3{\rm{bit/s/Hz}}$ with any one channel realization. Fig. \ref{fig:simulation4}(a) and Fig. \ref{fig:simulation4}(b) illustrate the harvested energy versus the updates of $t$ and iterations of variable $\left( {{{\bf{W}}_I}, {{\bf{W}}_E}, {\bf{V}}} \right)$, resp.. In Fig. \ref{fig:simulation4}(a), the curves from bottom to up represent the \textit{n}th iteration of $\left( {{{\bf{W}}_I}, {{\bf{W}}_E}, {\bf{V}}} \right)$ in turn. We can see that one circulation of the updates of $t$ is needed for every iteration of $\left( {{{\bf{W}}_I}, {{\bf{W}}_E}, {\bf{V}}} \right)$. Moreover, the harvested energy increases and converges to a fixed value finally. Also, to some degree, such a merit can be confirmed by Fig. \ref{fig:simulation4}(b). Similar to the curves in Fig. \ref{fig:simulation4}(a), different circles in Fig. \ref{fig:simulation4}(c) corresponding to different iteration of $\left( {{{\bf{W}}_I}, {{\bf{W}}_E}, {\bf{V}}} \right)$. From Fig. \ref{fig:simulation4}(c), we learn that $x_k^n \ge 0$ is satisfied and the value of barrier function $\frac{1}{t}\sum\limits_{k = 1}^K {\ln \left( {x_k^n} \right)} $ tends to zero with the update of $t$. Fig. \ref{fig:simulation4}(d) shows the innermost (GP method) iterations of Algorithm \ref{alg:Algorithm 3}. Since the iterations of GP are limited and the resulting solutions derived from the GP method are in analytical forms, a fast and efficient operation of our proposed algorithm can be achieved.

\begin{figure}[!t]
\centering
\includegraphics[width=3.2in]{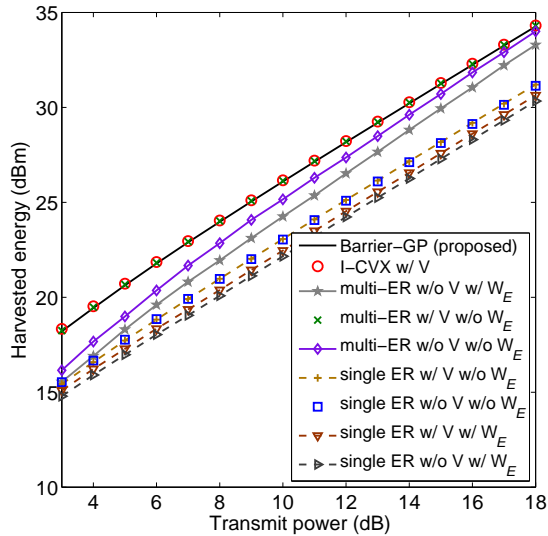}
\caption{Harvested energy versus the transmit power.}
\label{fig:simulation5}
\end{figure}

After the illustration of the convergence of our proposed algorithm, we will focus on the investigation of these questions. Where are the advantages of our proposed algorithm? Why do we introduce AN in the multiple ERs case rather than the single ER case? With these two questions, we carried on the following simulation tests. In Fig. \ref{fig:simulation5}, we demonstrate the energy harvesting performance versus the transmit power when fixing ${R_0} = 3{\rm{bit/s/Hz}}$. To test the performance of our proposed algorithm (marked as Barrier-GP (proposed) in the legend), we compare it with the I-CVX method with the introduction of AN   (marked as `I-CVX w/ ${\bf{V}}$'). From the plot, we can see that our proposed algorithm yields a consistent performance with the `I-CVX w/ ${\bf{V}}$' method which is deemed to be a common approach in achieving the optimal objective value. However, from Table \ref{table_2}, one can see that our proposed algorithm runs much faster than the `I-CVX w/ ${\bf{V}}$' method. The speedup is within the range of 55 to 400 times. Then, to test the benefit of AN in promoting energy harvesting, different scenarios, such as multiple ERs and single ER, with and without AN, with and without ${{\bf{W}}_E}$ cancellation at IR, etc., are taken into account in the plot. As an example, the legend `multi-ER w/o ${\bf{V}}$ w/ ${{\bf{W}}_E}$' represents the scenario of multiple ERs without the consideration of AN and the interference of energy signals is not cancelled at IR. Through the comparison of `I-CVX w/ ${\bf{V}}$' and `multi-ER w/o ${\bf{V}}$ w/ ${{\bf{W}}_E}$', we see a significant promotion in energy harvesting with the help of AN especially at the low power range. A similar result can also be affirmed from the comparison of `multi-ER w/ ${\bf{V}}$ w/o ${{\bf{W}}_E}$' and `multi-ER w/o ${\bf{V}}$ w/o ${{\bf{W}}_E}$'. To the single ER scenarios, it witnesses an identical performance between `single ER w/ ${\bf{V}}$ w/o ${{\bf{W}}_E}$' and `single ER w/o ${\bf{V}}$ w/o ${{\bf{W}}_E}$'. Though the `single ER w/ ${\bf{V}}$ w/ ${{\bf{W}}_E}$' case shows a better performance than `single ER w/o ${\bf{V}}$ w/ ${{\bf{W}}_E}$', the gain is almost negligible. To summarize, the above legends indicate that the AN plays a positive role in facilitating energy harvesting in the multiple ERs scenarios while not in the single ER scenarios.

\begin{table}[!t]
\caption{Average running time (in secs.) versus the transmit power}
\label{table_2}
\centering
\begin{tabular}{c|c|c|c|c|c}
\hline
\multirow{2}{*}{Method} & \multicolumn{5}{c}{Transmit Power (dB)}\\
\cline{2-6}
&3&6&9&12&15\\
\hline
Proposed&0.1036&0.1594&0.4741&0.9041&1.2387\\
I-CVX w/ ${\bf{V}}$&41.8818&44.7600&49.7995&55.0403&68.3917\\
\hline
\end{tabular}
\end{table}

\begin{figure}[!t]
\centering
\includegraphics[width=3.2in]{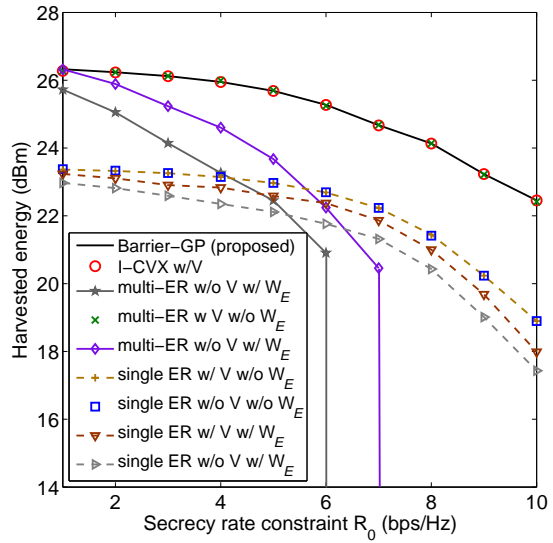}
\caption{Harvested energy versus the secrecy rate.}
\label{fig:simulation6}
\end{figure}

Fig. \ref{fig:simulation6} shows the harvested energy versus the secrecy rate when fixing  . The methods and scenarios considered here are same to Fig. \ref{fig:simulation5}. From Fig. \ref{fig:simulation6}, we have the following three observations: First, the harvested energy decreases as the secrecy rate grows up. This is due to the fact that more transmit power is assigned to fulfill the higher secrecy rate requirement. Hence, the power be used for energy transfer become less. Second, for the multiple ERs scenarios, the energy-rate region enlarges apparently under the aid of AN especially in the `w/ ${{\bf{W}}_E}$' case. Such an observation is consistent with our argument that the AN can both promote efficient energy transfer and guarantee secure communication in the multiple ERs MIMO system. Third, for the single ER scenarios, the energy-rate region changes slightly or even remain unchanged. This indicates the dispensability of AN in the single ER MIMO system.

\section{Conclusion}

This paper has studied the energy harvesting problems in secure MIMO systems with one or multiple ERs. For each scenario, two types of IR, which is able or not able to remove the interference from energy signals, have been taken into account. In the scenario of single ER, we studied two EHM problems (P1 and P2) and proposed an iterative WF-SVD algorithm which results in the semi-closed form solutions for problem P1. In the scenario of multiple ERs, we studied the AN-aided WS-EHM problem and proposed a barrier-GP based block GS algorithm with proved KKT point convergence of this problem. Numerical results demonstrated the superior performance of our proposed algorithms. In the future, we will consider to extend our attention to more complex and practical systems, e.g., multiple multi-antenna IRs, the channel state information between the transmitter and receivers are imperfect, the transmitter is equipped with a large number of antennas, etc.


%

\appendices
\section{Proof of Theorem \ref{thm:1}}

Simplify the expression of the Lagrangian in (\ref{eq:F14}) with given ${{\bf{W}}_{I0}}$, $\lambda $ and $\mu $, we have
\begin{equation}\label{eq:F41}
{\mathcal{L}}\left( {\lambda , \mu  {{\bf{W}}_I}} \right) = \lambda \ln \left( {{\bf{I}} + {{\bf{H}}^H}{{\bf{W}}_I}{\bf{H}}} \right) - {\rm{Tr}}\left( {{\bf{Q}}{{\bf{W}}_I}} \right) + {c_0},
\end{equation}
where ${c_0} = \lambda {\rm{Tr}}\left[ {{{\left( {{\bf{I}} + {{\bf{G}}^H}{{\bf{W}}_{I0}}{\bf{G}}} \right)}^{ - 1}}{{\bf{G}}^H}{{\bf{W}}_{I0}}{\bf{G}}} \right] + \mu P - \lambda \ln \left| {{\bf{I}} + {{\bf{G}}^H}{{\bf{W}}_{I0}}{\bf{G}}} \right| - \lambda {R_0}\ln 2$. Then, substitute ${{\bf{\tilde W}}_I}$ for ${{\bf{W}}_I}$ and let ${{\bf{\tilde W}}_I} = {{\bf{Q}}^{{1 \mathord{\left/{\vphantom {1 2}} \right.\kern-\nulldelimiterspace} 2}}}{{\bf{W}}_I}{{\bf{Q}}^{{1 \mathord{\left/{\vphantom {1 2}} \right.\kern-\nulldelimiterspace} 2}}}$, the equation (\ref{eq:F41}) can be reformed as
\begin{equation}\label{eq:F42}
\begin{array}{l}
{\mathcal{L}}\left( {\lambda , \mu  {{{\bf{\tilde W}}}_I}} \right) = \lambda \ln \left( {{\bf{I}} + {{\bf{H}}^H}{{\bf{Q}}^{ - {1 \mathord{\left/
 {\vphantom {1 2}} \right.
 \kern-\nulldelimiterspace} 2}}}{{{\bf{\tilde W}}}_I}{{\bf{Q}}^{ - {1 \mathord{\left/
 {\vphantom {1 2}} \right.
 \kern-\nulldelimiterspace} 2}}}^H{\bf{H}}} \right)\\
 ~~~~~~~~~~~~~~~~~~- {\rm{Tr}}\left( {{{{\bf{\tilde W}}}_I}} \right) + {c_0}.
\end{array}
\end{equation}
Do SVD about ${{\bf{H}}^H}{{\bf{Q}}^{ - {1 \mathord{\left/{\vphantom {1 2}} \right.\kern-\nulldelimiterspace} 2}}}$, i.e., ${{\bf{H}}^H}{{\bf{Q}}^{ - {1 \mathord{\left/{\vphantom {1 2}} \right.\kern-\nulldelimiterspace} 2}}} = {\bf{U}}{{\bf{\Lambda }}^{{1 \mathord{\left/{\vphantom {1 2}} \right.\kern-\nulldelimiterspace} 2}}}{{\bf{V}}^H}$, where ${\bf{U}} \in {\mathbb{C}^{{N_i} \times L}}$ and ${\bf{V}} \in {\mathbb{C}^{{N_t} \times L}}$, and substitute it into (\ref{eq:F42}). Subsequently, the optimal analytic solution for the maximization of (\ref{eq:F42}) can be achieved \cite{IEEEhowto:34}: ${{\bf{\tilde W}}_I} = {\bf{VD}}{{\bf{V}}^H}$, where ${\bf{D}} = {\rm{diag}}\left( {{p_1},{\kern 1pt} {\kern 1pt} {\kern 1pt} {p_2},{\kern 1pt} {\kern 1pt} {\kern 1pt}  \cdots ,{\kern 1pt} {\kern 1pt} {\kern 1pt} {p_L}} \right)$. Now, let us seek for the solution of the diagonal element ${p_i}, i = 1,  \cdots , L$. Substituting ${{\bf{\tilde W}}_I} = {\bf{VD}}{{\bf{V}}^H}$ into (\ref{eq:F42}), we have
\begin{equation}\label{eq:F43}
{\mathcal{L}}(\lambda , \mu , {\bf{D}}) = \lambda \ln \left( {{\bf{I}} + {\bf{D\Lambda }}} \right) - {\rm{Tr}}\left( {\bf{D}} \right) + {c_0}.
\end{equation}
Maximizing (\ref{eq:F43}) about ${\bf{D}}$, combine with the standard WF power allocation method \cite{IEEEhowto:34}, the solution ${p_i}$ can be obtained: ${p_i} = \left( {\lambda  - \frac{1}{{\delta _i^2}}} \right),{\kern 1pt} {\kern 1pt} {\kern 1pt} {\kern 1pt} i = 1,{\kern 1pt} {\kern 1pt} {\kern 1pt}  \cdots ,{\kern 1pt} {\kern 1pt} {\kern 1pt} L$. Thus, ${{\bf{\tilde W}}_I} = {{\bf{Q}}^{{1 \mathord{\left/
 {\vphantom {1 2}} \right.\kern-\nulldelimiterspace} 2}}}{{\bf{W}}_I}{{\bf{Q}}^{{1 \mathord{\left/
 {\vphantom {1 2}} \right.\kern-\nulldelimiterspace} 2}}}\\ = {\bf{VD}}{{\bf{V}}^H}$ implies ${\bf{W}}_{_I}^* = {{\bf{Q}}^{ - {1 \mathord{\left/ {\vphantom {1 2}} \right. \kern-\nulldelimiterspace} 2}}}{\bf{VD}}{{\bf{V}}^H}{{\bf{Q}}^{ - {1 \mathord{\left/ {\vphantom {1 2}} \right.\kern-\nulldelimiterspace} 2}}}$, and the diagonal matrix ${\bf{D}} = {\rm{diag}}\left( {{p_1},{\kern 1pt} {\kern 1pt} {\kern 1pt} {\kern 1pt} {\kern 1pt} {\kern 1pt}  \cdots ,{\kern 1pt} {\kern 1pt} {\kern 1pt} {p_L}} \right)$ is established with given ${p_i}, i = 1,  \cdots , L$.

Given ${{\bf{W}}_I}$, the dual variables $\lambda$ and $\mu$ can be optimized by minimizing the dual function (\ref{eq:F14b}). Here, the subgradient-based approach, e.g., ellipsoid method \cite{IEEEhowto:35}, is employed to deal with the dual problem $\left( {\widetilde {\textsc{P1}} - 1 - D} \right)$. According to such a method, the subgradient, a key index during operations, of $d\left( {\lambda , \mu } \right)$ at point $\left[ {\lambda , \mu } \right]$ can be defined as $\left[ {\tilde C_{_{s1}}^* - {C_0}, {\kern 1pt} \alpha P - {\rm{Tr}}\left( {{\bf{W}}_{_I}^*} \right)} \right]$. Once the optimal dual solutions ${\lambda ^*}$ and ${\mu ^*}$ are figured out, subsequently, the optimal diagonal element $p_{_i}^* = \left( {{\lambda ^*} - \frac{1}{{\delta _i^2}}} \right),{\kern 1pt} {\kern 1pt} {\kern 1pt} {\kern 1pt} i = 1,{\kern 1pt} {\kern 1pt} {\kern 1pt}  \cdots ,{\kern 1pt} {\kern 1pt} {\kern 1pt} L$ will be derived as well as the optimal transmit information covariance ${\bf{W}}_{_I}^* = {{\bf{Q}}^{ - {1 \mathord{\left/ {\vphantom {1 2}} \right. \kern-\nulldelimiterspace} 2}}}{\bf{V}}{{\bf{D}}^*}{{\bf{V}}^H}{{\bf{Q}}^{ - {1 \mathord{\left/ {\vphantom {1 2}} \right.
 \kern-\nulldelimiterspace} 2}}}$ which converges to the optimal solution to subproblem $\left( {\widetilde {\textsc{P1}} - 1} \right)$. Thus, the proof of Theorem \ref{thm:1} is finished.

\section{Proof of Proposition \ref{pro:1}}

To have a better understanding of the update processes of the Lagrangian multipliers $\lambda $ and $\mu $ by applying the subgradient-based approach, we need the following lemma extracted from \cite{IEEEhowto:35}.
\newtheorem{Lem}{\textit{\underline{Lemma}}}[section]
\begin{Lemma}\label{lem:2}
Considering the problem of maximizing a convex function $f$: ${\mathbb{R}^n} \to \mathbb{R} $ and an ellipsoid ${\bm{{\rm O}}}$ be descript as
\[{\bm{{\rm O}}} = \left\{ {{\bf{z}}|{{\left( {{\bf{z}} - {\bf{x}}} \right)}^T}{{\bf{P}}^{ - 1}}\left( {{\bf{z}} - {\bf{x}}} \right) \le 1} \right\},\]
where ${\bf{P}} \in {\bf{S}}_{ +  + }^n$ gives the size and shape of ${\bm{{\rm O}}}$; ${\bf{x}} \in {\mathbb{R}^n}$, each element inside is nonnegative, is the center of the ellipsoid ${\bm{{\rm O}}}$. Computing the subgradient of the objective function $f$ at the \textit{j}th iteration, it has ${{\bf{s}}^j} = \partial f\left( {{{\bf{x}}^j}} \right)$. The center point ${\bf{x}}$ is updated as ${{\bf{x}}^j} = {{\bf{x}}^{j - 1}} - \frac{1}{{n + 1}}{{\bf{P}}^{j - 1}}{\bf{\tilde s}}$, where ${\bf{\tilde s}} = \frac{{{{\bf{s}}^{j - 1}}}}{{\sqrt {{{\left( {{{\bf{s}}^{j - 1}}} \right)}^T}{{\bf{P}}^{j - 1}}{{\bf{s}}^{j - 1}}} }}$ and ${{\bf{P}}^j} = \frac{{{n^2}}}{{{n^2} - 1}}\left( {{{\bf{P}}^{j - 1}} - \frac{2}{{n + 1}}{{\bf{P}}^{j - 1}}{\bf{\tilde s}}{{{\bf{\tilde s}}}^T}{{\bf{P}}^{j - 1}}} \right)$. Stopping the iteration once $\sqrt {{{\left( {{{\bf{s}}^j}} \right)}^T}{{\bf{P}}^j}{{\bf{s}}^j}}  \le \varepsilon $, where $\varepsilon  > 0$, fulfills.
\end{Lemma}

According to Lemma \ref{lem:2}, using the ellipsoid method, we assume that point ${\bf{x}} = {\left[ {\lambda , \mu } \right]^T}$ is the center of the ellipsoid ${\bm{{\rm O}}}$. Then, given the initial matrix ${{\bf{A}}^0} \in {\bf{S}}_{ +  + }^2$ which characterize the size and the orientation of the ellipsoid ${\bm{{\rm O}}}$, the center point ${\left[ {\lambda , \mu } \right]^T}$ can be updated as ${\left[ {{\lambda ^j},{\kern 1pt} {\kern 1pt} {\mu ^j}} \right]^T} = {\left[ {{\lambda ^{j - 1}},{\kern 1pt} {\kern 1pt} {\mu ^{j - 1}}} \right]^T} - \frac{1}{3}{{\bf{A}}^{j - 1}}{\bf{\tilde s}}$, where ${\bf{\tilde s}} = \frac{{\bf{s}}}{{\sqrt {{{\bf{s}}^T}{{\bf{A}}^{j - 1}}{\bf{s}}} }}$ and that ${\bf{s}} = {\left[ {\frac{{\partial \mathcal{L}\left( {{\bf{W}}_I^*,{\kern 1pt} {\kern 1pt} {\lambda ^{j - 1}},{\kern 1pt} {\kern 1pt} {\mu ^{j - 1}}} \right)}}{{\partial {\lambda ^{j - 1}}}},{\kern 1pt} {\kern 1pt} \frac{{\partial \mathcal{L}\left( {{\bf{W}}_I^*,{\kern 1pt} {\kern 1pt} {\lambda ^{j - 1}},{\kern 1pt} {\kern 1pt} {\mu ^{j - 1}}} \right)}}{{\partial {\mu ^{j - 1}}}}} \right]^T} = {\left[ {\tilde C_{_{s1}}^* - {C_0}, {\kern 1pt} \alpha P - {\rm{Tr}}\left( {{\bf{W}}_{_I}^*} \right)} \right]^T}$, ${\bf{W}}_I^*$ is the corresponding optimal value obtained from the \textit{j}th iteration. The matrix ${\bf{A}}$ can be updated as: ${{\bf{A}}^j} = \frac{4}{3}\left( {{{\bf{A}}^{j - 1}} - \frac{2}{3}{{\bf{A}}^{j - 1}}{\bf{\tilde s\tilde s}}{{\bf{A}}^{j - 1}}} \right)$. Moreover, the stopping criterion is thus formed as ${\left( {{{\bf{s}}^T}{{\bf{A}}^j}{\bf{s}}} \right)^{{1 \mathord{\left/{\vphantom {1 2}} \right.\kern-\nulldelimiterspace} 2}}} \le \varepsilon $.

Next, let’s discuss the convergence of the iteration. From the volume formula $V\left( {\bm{{\rm O}}} \right){\rm{ = }}{\beta _n}\sqrt {\det \left( {\bf{P}} \right)} $ in which ${\beta _n} = {\raise0.7ex\hbox{${{\pi ^{{n \mathord{\left/
 {\vphantom {n 2}} \right.
 \kern-\nulldelimiterspace} 2}}}}$} \!\mathord{\left/
 {\vphantom {{{\pi ^{{n \mathord{\left/
 {\vphantom {n 2}} \right.
 \kern-\nulldelimiterspace} 2}}}} {\Gamma \left( {{n \mathord{\left/
 {\vphantom {n 2}} \right.
 \kern-\nulldelimiterspace} 2} + 1} \right)}}}\right.\kern-\nulldelimiterspace}
\!\lower0.7ex\hbox{${\Gamma \left( {{n \mathord{\left/
 {\vphantom {n 2}} \right.
 \kern-\nulldelimiterspace} 2} + 1} \right)}$}}$, we elicit the volume whose size and shape are determined by matrix ${\bf{A}} \in {\bf{S}}_{ +  + }^2$: $V\left( {\bm{{\rm O}}} \right){\rm{ = }}{\beta _2}\sqrt {\det \left( {\bf{A}} \right)} $. Suppose that our original ellipsoid ${{\bm{{\rm O}}}^0}$ contains the minimizer in its interior, i.e., ${\left[ {{\lambda ^*}, {\mu ^*}} \right]^T} \in {{\bm{{\rm O}}}^0}$, the main idea of the ellipsoid method is to shrink the volume of the resulting ellipsoid of previous iteration such that ${\mathcal{L}_{best}}\left( {{{\bf{W}}^*}, {\lambda ^j}, {\mu ^j}} \right) \le \mathcal{L}\left( {{{\bf{W}}^*}, {\lambda ^*}, {\mu ^*}} \right) + \varepsilon $ satisfies, where ${\left[ {{\lambda ^j}, {\mu ^j}} \right]^T}$ is the center point of the ellipsoid ${{\bm{{\rm O}}}^j}$ at the \textit{j}th iteration. Therefore, the convergence can be judged based on the change of the volume of ellipsoid ${{\bm{{\rm O}}}^k}$ which can be given as:
\begin{equation}\label{eq:F44}
\begin{array}{l}
V\left( {{{\bm{{\rm O}}}^j}} \right) = {\beta _2}\sqrt {\det \left( {{{\bf{A}}^j}} \right)} \\
    \;~~~~~~~~= {\beta _2}{\left[ {\det \left( {\frac{4}{3}\left( {{{\bf{A}}^{j - 1}} - \frac{2}{3}{{\bf{A}}^{j - 1}}{\bf{\tilde s}}{{{\bf{\tilde s}}}^T}{{\bf{A}}^{j - 1}}} \right)} \right)} \right]^{\frac{1}{2}}}\\
\; {\kern 1pt} {\kern 1pt} {\kern 1pt} {\kern 1pt} {\kern 1pt} {\kern 1pt} {\kern 1pt} {\kern 1pt} {\kern 1pt} {\kern 1pt} {\kern 1pt} {\kern 1pt} {\kern 1pt} {\kern 1pt} {\kern 1pt} {\kern 1pt} {\kern 1pt} {\kern 1pt} {\kern 1pt} {\kern 1pt} {\kern 1pt} {\kern 1pt} {\kern 1pt} {\kern 1pt} {\kern 1pt} {\kern 1pt} {\kern 1pt} {\kern 1pt}  = {\beta _2}{\left[ {\det \left( {\frac{4}{3}\left( {{{\bf{A}}^{j - 1}} - \frac{2}{3}{{\bf{A}}^{j - 1}}\frac{{{\bf{s}}{{\bf{s}}^T}}}{{{{\bf{s}}^T}{{\bf{A}}^{j - 1}}{\bf{s}}}}{{\bf{A}}^{j - 1}}} \right)} \right)} \right]^{\frac{1}{2}}}\\
{\kern 1pt} {\kern 1pt} {\kern 1pt} {\kern 1pt} {\kern 1pt} {\kern 1pt} {\kern 1pt} {\kern 1pt} {\kern 1pt} {\kern 1pt} {\kern 1pt} {\kern 1pt} {\kern 1pt} {\kern 1pt} {\kern 1pt} {\kern 1pt} {\kern 1pt} {\kern 1pt} {\kern 1pt} {\kern 1pt} {\kern 1pt} {\kern 1pt} {\kern 1pt} {\kern 1pt} {\kern 1pt} {\kern 1pt} {\kern 1pt} {\kern 1pt} {\kern 1pt} {\kern 1pt} {\kern 1pt} \mathop  = \limits^{(a)} \frac{4}{3}{\beta _2}{\left[ {\det ({{\bf{A}}^{j - 1}})} \right]^{\frac{1}{2}}}{\left[ {\det \left( {{\bf{I}} - \frac{2}{3}\frac{{{\bf{s}}{{\bf{s}}^T}}}{{{{\bf{s}}^T}{{\bf{A}}^{j - 1}}{\bf{s}}}}{{\bf{A}}^{j - 1}}} \right)} \right]^{\frac{1}{2}}}\\
~{\kern 1pt} {\kern 1pt} {\kern 1pt} {\kern 1pt} {\kern 1pt} {\kern 1pt} {\kern 1pt} {\kern 1pt} {\kern 1pt} {\kern 1pt} {\kern 1pt} {\kern 1pt} {\kern 1pt} {\kern 1pt} {\kern 1pt} {\kern 1pt} {\kern 1pt} {\kern 1pt} {\kern 1pt} {\kern 1pt} {\kern 1pt} {\kern 1pt} {\kern 1pt} {\kern 1pt} {\kern 1pt} {\kern 1pt} {\kern 1pt} {\kern 1pt} \mathop  = \limits^{(b)} \frac{4}{3}{\beta _2}{\left[ {\det \left( {{{\bf{A}}^{j - 1}}} \right)} \right]^{\frac{1}{2}}}{\left[ {1 - {\rm{Tr}}\left( {\frac{2}{3}\frac{{{\bf{s}}{{\bf{s}}^T}}}{{{{\bf{s}}^T}{{\bf{A}}^{j - 1}}{\bf{s}}}}{{\bf{A}}^{j - 1}}} \right)} \right]^{\frac{1}{2}}}\\
     ~~~~~~~~~ = \frac{4}{3}{\beta _2}{\left[ {\det \left( {{{\bf{A}}^{j - 1}}} \right)} \right]^{\frac{1}{2}}}{\left[ {1 - \frac{2}{3}\frac{{{\rm{Tr}}\left( {{\bf{s}}{{\bf{s}}^T}{{\bf{A}}^{j - 1}}} \right)}}{{{{\bf{s}}^T}{{\bf{A}}^{j - 1}}{\bf{s}}}}} \right]^{\frac{1}{2}}}\\
     ~~~~~~~~~ = \frac{4}{3}{\left[ {1 - \frac{2}{3}} \right]^{\frac{1}{2}}}{\beta _2}{\left[ {\det \left( {{{\bf{A}}^{j - 1}}} \right)} \right]^{\frac{1}{2}}}\\
     ~~~~~~~~~ {\kern 1pt} \mathop  < \limits^{(c)} {e^{{{ - 1} \mathord{\left/
 {\vphantom {{ - 1} 4}} \right.
 \kern-\nulldelimiterspace} 4}}}V\left( {{{\bm{{\rm O}}}^{j - 1}}} \right)\\
     ~~~~~~~~~ {\kern 1pt} \mathop  < \limits^{(d)} {e^{{{ - J} \mathord{\left/
 {\vphantom {{ - J} 4}} \right.
 \kern-\nulldelimiterspace} 4}}}V\left( {{{\bm{{\rm O}}}^0}} \right),
\end{array}
\end{equation}
where $\left( a \right)$ is due to the fact that $\det ({\bf{XY}}) = \det ({\bf{X}})\det ({\bf{Y}})$; $\left( b \right)$ is due to the fact that $\det \left( {{\bf{I}} - {\bf{A}}} \right) = 1 - {\rm{Tr}}\left( {\bf{A}} \right)$ when ${\rm{rank}}\left( {\bf{A}} \right) \le 1$, here, ${\rm{rank}}\left( {{\bf{s}}{{\bf{s}}^T}} \right) = 1$ and ${{\bf{A}}^{j - 1}} \in {\bf{S}}_{ +  + }^2$ lead to ${\rm{rank}}\left( {{\bf{s}}{{\bf{s}}^T}{{\bf{A}}^{j - 1}}} \right) \le 1$; $\left( c \right)$ is result from a conclusion in \cite{IEEEhowto:35} that ${\left( {\frac{n}{{n + 1}}} \right)^{{{(n + 1)} \mathord{\left/
 {\vphantom {{(n + 1)} 2}} \right.
 \kern-\nulldelimiterspace} 2}}}{\left( {\frac{n}{{n - 1}}} \right)^{{{(n - 1)} \mathord{\left/
 {\vphantom {{(n - 1)} 2}} \right.
 \kern-\nulldelimiterspace} 2}}} < {e^{{{ - 1} \mathord{\left/
 {\vphantom {{ - 1} {2n}}} \right.
 \kern-\nulldelimiterspace} {2n}}}}$ and $\left( d \right)$ is a recursive conclusion based on the derivations above it. From (\ref{eq:F44}), we confirm that the volume of the ellipsoid ${{\bm{{\rm O}}}^j}$ decreases after each iteration. Therefore, the next work we need to do is to find the upper bound of our iterations. To begin with, let’s define a ball
\begin{equation}\label{eq:F45}
{\bm{{\rm B}}} = \left\{ {{\bf{z}}|\left\| {{\bf{z}} - {{\bf{x}}^*}} \right\| \le {\raise0.7ex\hbox{$\varepsilon $} \!\mathord{\left/
 {\vphantom {\varepsilon  {{l_s}}}}\right.\kern-\nulldelimiterspace}
\!\lower0.7ex\hbox{${{l_s}}$}}} \right\},
\end{equation}
where ${l_s} = \mathop {\max }\limits_{{\bf{s}} \in \partial \mathcal{L}\left( {{\bf{W}},\lambda ,\mu } \right)} \;\left\| {\bf{s}} \right\|$ is the maximum length of the subgradients over the initial ellipsoid ${{\bm{{\rm O}}}^0}$ and ${\bf{B}} \subseteq {{\bm{{\rm O}}}^0}$. With this, we have $\mathcal{L}\left( {{\bf{W}},{\kern 1pt} {\bf{z}}} \right) \le \mathcal{L}\left( {{\bf{W}},{\kern 1pt} {\kern 1pt} {{\bf{x}}^*}} \right) + \varepsilon $ when ${\bf{z}} \in {\bf{B}}$, while for ${\bf{z}} \notin {\bf{B}}$, it is deemed that $\mathcal{L}\left( {{\bf{W}},{\kern 1pt} {\bf{z}}} \right) \ge \mathcal{L}\left( {{\bf{W}},{\kern 1pt} {\kern 1pt} {{\bf{x}}^*}} \right) + \varepsilon $. Since ${{\bf{x}}^*}$ is defined as the final solution of the iteration, i.e., the achieved optimal solution, and ball ${\bm{{\rm B}}}$ is the infinitesimal region contains ${{\bf{x}}^*}$, we have ${\bf{B}} \subseteq {{\bm{{\rm O}}}^j}$ in iterations $1,{\kern 1pt}  \cdots ,{\kern 1pt} {\kern 1pt} J$. Hence, $V\left( {{{\bm{{\rm O}}}^j}} \right) \ge V\left( {\bf{B}} \right)$, which leads to
\begin{equation}\label{eq:F46}
{e^{{{ - J} \mathord{\left/
 {\vphantom {{ - J} 4}} \right.
 \kern-\nulldelimiterspace} 4}}}V\left( {{{\bm{{\rm O}}}^0}} \right) \ge V\left( {\bf{B}} \right) = {\beta _2}{\left( {{\raise0.7ex\hbox{$\varepsilon $} \!\mathord{\left/
 {\vphantom {\varepsilon  {{l_s}}}}\right.\kern-\nulldelimiterspace}
\!\lower0.7ex\hbox{${{l_s}}$}}} \right)^2}.
\end{equation}
Taking the initial ellipsoid ${{\bm{{\rm O}}}^0}$ to be a ball, i.e., ${{\bf{x}}^0} = {\left[ {{\lambda ^0}, {\mu ^0}} \right]^T} = {\bf{0}}$ and ${{\bf{A}}^0} = {r^2}{\bf{I}}$. Then, do log operations of the first and last terms in (\ref{eq:F46}), we have
\begin{equation}\label{eq:F47}
 - \frac{J}{4} + 2\ln r \ge 2\ln \frac{\varepsilon }{{{l_s}}},
\end{equation}
where $V\left( {{{\bm{{\rm O}}}^0}} \right) = {\beta _2}{r^2}$. Finally, the maximum iteration number can be expressed as
\begin{equation}\label{eq:F48}
J \le 8\ln \frac{{r{l_s}}}{\varepsilon }.
\end{equation}
So, we conclude that it takes no more than $8\ln \frac{{r{l_s}}}{\varepsilon }$ iterations to achieve $\mathcal{L}\left( {{\bf{W}}, {\lambda ^*}, {\mu ^*}} \right)$ with error at most $\varepsilon $.

\section{Proof of Proposition \ref{pro:4}}

Before the start of the proof, we need construct an optimization problem based on the subproblems $\left( {\widetilde {\textsc{P3}} - 1} \right)$, $\left( {\widetilde {\textsc{P3}} - 2} \right)$ and $\left( {\widetilde {\textsc{P3}} - 3\rm{A}} \right)$ solved in each iteration. Such a problem is formed as
\begin{subequations}\label{eq:F49}
\begin{equation}\label{eq:F49a}
\left( {\overline {{\rm{P}}3} } \right):         \mathop {\max }\limits_{{{\bf{W}}_I}, {{\bf{W}}_E},{\kern 1pt} {\kern 1pt} {\bf{V}}}  {{\rm{E}}_{{\rm{WS}}}}\left( {{{\bf{W}}_I}, {{\bf{W}}_E}, {\bf{V}}} \right) + \frac{1}{t}\sum\limits_{k = 1}^K {\ln \left( {{x_k}} \right)}
\end{equation}
\begin{equation}\label{eq:F49b}
s.t. ~~~~~~{\rm{Tr}}\left( {{{\bf{W}}_I} + {{\bf{W}}_E} + {\bf{V}}} \right) \le P,
\end{equation}
\begin{equation}\label{eq:F49c}
~~~~~{{\bf{W}}_I} \succeq {\bf{0}}, {{\bf{W}}_E} \succeq {\bf{0}}, {\bf{V}} \succeq {\bf{0}}, {{\bf{{\rm T}}}_0} \succeq {\bf{0}}, {{\bf{{\rm T}}}_k} \succeq {\bf{0}}, \forall k,
\end{equation}
\end{subequations}
where ${x_k} = {x_k}\left( {{{\bf{W}}_I}, {{\bf{W}}_E}, {\bf{V}}, \{ {{{\bf{T}}_k}} \}_{k = 0}^K} \right) = \mathop {\max }\limits_{{{\bf{{\rm T}}}_0} \succeq {\bf{0}}, {{\bf{{\rm T}}}_k} \succeq {\bf{0}}} {\theta _I}(\\ {{{\bf{W}}_I}, {{\bf{W}}_E}, {\bf{V}}, {{\bf{T}}_0}}) - {\theta _{E,k}}\left( {{{\bf{W}}_I}, {{\bf{W}}_E}, {\bf{V}}, {{\bf{T}}_k}} \right) - {R_0}\ln 2$.
Next, based on this, we complete the proof through two steps. First, we demonstrate that every limit point generate by the block GS method (cf. Algorithm \ref{alg:Algorithm 3}) is a critical point of problem $\left( {\overline {\textsc{P3}} } \right)$. Then, we show that the critical point of problem $\left( {\overline {\textsc{P3}} } \right)$ is also a critical point of the original problem $\left( {{\rm{P}}3} \right)$. Only then can our statement in Proposition \ref{pro:4} be ascertained.

\textit{In the first step}, we first introduce the following lemma to support our preliminary convergence result.
\newtheorem{Lemm}{\textit{\underline{Lemma}}}[section]
\begin{Lemma}\label{lem:3}(\cite{IEEEhowto:30}, Proposition 6)
Consider such a problem:
\begin{equation}\label{eq:F50}
 \begin{array}{l}
{\kern 1pt} {\kern 1pt} {\kern 1pt} {\kern 1pt} {\kern 1pt} ~~\max ~ f\left( {\bf{x}} \right)\\
s.t.{\kern 1pt} {\kern 1pt} {\kern 1pt} {\kern 1pt} {\kern 1pt} {\kern 1pt} {\kern 1pt} {\bf{x}} \in {\bf{X}} = {{\bf{X}}_1} \times {{\bf{X}}_2} \times  \cdots  \times {{\bf{X}}_m} \subseteq {\mathbb{R}^n},{\kern 1pt} {\kern 1pt} \left( {m > 2} \right)
\end{array}
\end{equation}
which can be solved iteratively by m-block GS method. Suppose that $f:{\kern 1pt} {\kern 1pt} {\kern 1pt} {\mathbb{R}^n} \to \mathbb{R}$ is pseudoconcave on ${\bf{X}}$ and the level set $\ell _{\bf{X}}^0: = \left\{ {{\bf{x}} \in {\bf{X}}:f\left( {\bf{x}} \right) \ge f\left( {{{\bf{x}}^0}} \right),{{\bf{x}}^0} \in {\bf{X}}} \right\}$ is compact. Then, it can be confirmed that the sequence $\left\{ {{{\bf{x}}^n}} \right\}$ produced by the m-block GS method has limit points and every limit point of $\left\{ {{{\bf{x}}^n}} \right\}$ is a critical point of (\ref{eq:F50}).
\end{Lemma}

It is easy to see that the objective (\ref{eq:F49a}) is concave w.r.t. each variable, fixing the others. And, the level set $\ell _{\bf{X}}^0$ is compact for every $\left( {{\bf{W}}_I^0,{\kern 1pt} {\kern 1pt} {\bf{W}}_E^0,{{\bf{V}}^0},{\bf{T}}_0^0,\left\{ {{\bf{T}}_k^0} \right\}_{k = 1}^K} \right) \in \bm{\aleph} $, of which $\bm{\aleph}  = \left\{ {{{\bf{W}}_I},{\kern 1pt} {\kern 1pt} {{\bf{W}}_E},{\bf{V}},{{\bf{T}}_0},\left\{ {{{\bf{T}}_k}} \right\}_{k = 1}^K|\;{x_k} \ge 0,{\kern 1pt} {\kern 1pt} (\ref{eq:F49b}),(\ref{eq:F49c})} \right\}$, on account of that the iterates $\left( {{\bf{W}}_{_I}^n,{\kern 1pt} {\kern 1pt} {\bf{W}}_{_E}^n,{{\bf{V}}^n},{\bf{T}}_{_0}^n,\left\{ {{\bf{T}}_{_k}^n} \right\}_{k = 1}^K} \right)$ are bounded (own to the total transmit power constraint) \cite{IEEEhowto:30}. In consequence, invoking Lemma \ref{lem:3}, we conclude that $\left( {{\bf{W}}_{_I}^n,{\kern 1pt} {\kern 1pt} {\bf{W}}_{_E}^n,{{\bf{V}}^n},{\bf{T}}_{_0}^n,\left\{ {{\bf{T}}_{_k}^n} \right\}_{k = 1}^K} \right)$ have limit points and every limit point produced by the block GS method is a critical point of problem $\left( {\overline {\textsc{P3}} } \right)$.

\textit{In the second step}, our endeavor is to prove that the critical point of problem $\left( {\overline {\textsc{P3}} } \right)$ is also a critical point of our original problem $\left( {{\rm{P}}3} \right)$. Suppose that   is a critical point of problem $\left( {{\rm{P}}3} \right)$. The Lagrangian of problem $\left( {\overline {\textsc{P3}} } \right)$ can be written as
\[\begin{array}{l}
\mathcal{L}\left( {\bm{\chi }} \right){\rm{ = }}{{\rm{E}}_{{\rm{WS}}}}\left( {{{\bf{W}}_I}, {{\bf{W}}_E}, {\bf{V}}} \right) + \frac{1}{t}\sum\limits_{k = 1}^K {\ln \left( {{x_k}} \right)} \\
 ~~~~~~~~- \upsilon \left[ {{\rm{Tr}}\left( {{{\bf{W}}_I} + {{\bf{W}}_E} + {\bf{V}}} \right) - P} \right] + {\rm{Tr}}\left( {{{\bf{\Phi }}_{{{\bf{W}}_I}}}{{\bf{W}}_I}} \right)\\
 ~~~~~~~~+ {\rm{Tr}}\left( {{{\bf{\Phi }}_{{{\bf{W}}_E}}}{{\bf{W}}_E}} \right) + {\rm{Tr}}\left( {{{\bf{\Phi }}_{\bf{V}}}{\bf{V}}} \right) + {\rm{Tr}}\left( {{{\bf{\Phi }}_{{{\bf{T}}_0}}}{{\bf{T}}_0}} \right)\\
 ~~~~~~~~+ \sum\limits_{k = 1}^K {{\rm{Tr}}\left( {{{\bf{\Phi }}_{{{\bf{T}}_k}}}{{\bf{T}}_k}} \right)} ,
\end{array}\]
where ${\bm{\chi }}$ denotes a collection of all the primal and dual variables of problem $\left( {\overline {\textsc{P3}} } \right)$, $\upsilon  \in {\mathbb{R}_ + }$, ${{\bf{\Phi }}_{{{\bf{W}}_I}}} \in {\mathbb{H}}_ + ^{{N_t}}$, ${{\bf{\Phi }}_{{{\bf{W}}_E}}} \in {\mathbb{H}}_ + ^{{N_t}}$, ${{\bf{\Phi }}_{\bf{V}}} \in {\mathbb{H}}_ + ^{{N_t}}$, ${{\bf{\Phi }}_{{{\bf{T}}_0}}} \in {\mathbb{H}}_ + ^{{N_i}}$ and ${{\bf{\Phi }}_{{{\bf{T}}_k}}} \in {\mathbb{H}}_ + ^{{N_e}}$ are the dual variables associated with (\ref{eq:F49b}), ${{\bf{W}}_I} \succeq {\bf{0}}$, ${{\bf{W}}_E} \succeq {\bf{0}}$, ${\bf{V}} \succeq {\bf{0}}$, ${{\bf{T}}_0} \succeq {\bf{0}}$ and ${{\bf{T}}_k} \succeq {\bf{0}}$, resp.. Since $\left( {{\bf{W}}_{_I}^{\rm{*}},{\kern 1pt} {\kern 1pt} {\bf{W}}_{_E}^{\rm{*}},{{\bf{V}}^{\rm{*}}}} \right)$ maximizes $\mathcal{L}\left( {\bm{\chi }} \right)$ for given ${\bf{T}}_0^*$ and $\left\{ {{\bf{T}}_k^*} \right\}_{k = 1}^K$, its gradients w.r.t. ${{\bf{W}}_I}$, ${{\bf{W}}_E}$ and ${\bf{V}}$ vanishes at $\left( {{\bf{W}}_{_I}^{\rm{*}},{\kern 1pt} {\kern 1pt} {\bf{W}}_{_E}^{\rm{*}},{{\bf{V}}^{\rm{*}}}} \right)$, see (\ref{eq:F51})
\begin{figure*}[!t]
\normalsize
\begin{equation}\label{eq:F51}
\left\{ \begin{array}{l}
{\nabla _{{{\bf{W}}_I}}}\mathcal{L}\left( {\bm{\chi }} \right) = \sum\limits_{k = 1}^K {\sigma _{_{e,k}}^2{\mu _k}{\eta _k}{{\bf{G}}_k}{\bf{G}}_k^H}  + \frac{1}{t}\sum\limits_{k = 1}^K {\frac{1}{{x_{_k}^*}}\underbrace {\left( {{{\bf{\Omega }}^*} - {{\bf{G}}_k}{\bf{T}}_k^*{\bf{G}}_k^H} \right)}_{{\nabla _{{{\bf{W}}_I}}}x_{_k}^*}}  - {\upsilon ^*}{\bf{I}} + {\bf{\Phi }}_{{{\bf{W}}_I}}^* = {\bf{0}},\\
{\nabla _{{{\bf{W}}_E}}}\mathcal{L}\left( {\bm{\chi }} \right) = \sum\limits_{k = 1}^K {\sigma _{_{e,k}}^2{\mu _k}{\eta _k}{{\bf{G}}_k}{\bf{G}}_k^H}  + \frac{1}{t}\sum\limits_{k = 1}^K {\frac{1}{{x_{_k}^*}}\underbrace {\left( { - {\bf{HT}}_0^*{{\bf{H}}^H} + {{\bf{\Omega }}^*}} \right)}_{{\nabla _{{{\bf{W}}_E}}}x_{_k}^*} - {\upsilon ^*}{\bf{I}} + {\bf{\Phi }}_{{{\bf{W}}_E}}^* = {\bf{0}},} \\
{\nabla _{{{\bf{W}}_E}}}\mathcal{L}\left( {\bm{\chi }} \right) = \sum\limits_{k = 1}^K {\sigma _{_{e,k}}^2{\mu _k}{\eta _k}{{\bf{G}}_k}{\bf{G}}_k^H}  + \frac{1}{t}\sum\limits_{k = 1}^K {\frac{1}{{x_{_k}^*}}\underbrace {\left( { - {\bf{HT}}_0^*{{\bf{H}}^H} + {{\bf{\Omega }}^*} + {{\bf{\Psi }}^*} - {{\bf{G}}_k}{\bf{T}}_k^*{\bf{G}}_k^H} \right)}_{{\nabla _{\bf{V}}}x_{_k}^*} - {\upsilon ^*}{\bf{I}} + {\bf{\Phi }}_{\bf{V}}^* = {\bf{0}},}
\end{array} \right.
\end{equation}
\hrulefill
\vspace*{4pt}
\end{figure*}
where $x_{_k}^* = {x_k}\left( {{\bf{W}}_{_I}^*, {\bf{W}}_{_E}^*, {{\bf{V}}^*}, \left\{ {{\bf{T}}_{_k}^*} \right\}_{k = 0}^K} \right)$, ${{\bf{\Omega }}^{\rm{*}}} = {\bf{H}}[ {\bf{I}} + {{\bf{H}}^H}( {\bf{W}}_I^{\rm{*}} + \\ {\bf{W}}_E^{\rm{*}} + {{\bf{V}}^{\rm{*}}} ){\bf{H}} ]^{ - 1}{{\bf{H}}^H}$ and ${{\bf{\Psi }}^{\rm{*}}} = {{\bf{G}}_k}[ {\bf{I}} + {\bf{G}}_k^H{{\bf{V}}^{\rm{*}}}{{\bf{G}}_k} ]^{ - 1}{\bf{G}}_k^H$. Defining
\[\lambda _k^*\left( t \right) = \frac{1}{{tx_k^*}}, \forall k\]
we claim that $\lambda _k^*\left( t \right)$ is the dual variable w.r.t. the \textit{k}th secrecy rate constraint in problem $\left( {{\rm{P}}3} \right)$ and know that $\lambda _k^*\left( t \right) > 0$ due to $x_k^* > 0, \forall k$. Moreover, as $t \to \infty $, we have $\lambda _k^*\left( t \right)x_k^* = 0, \forall k$. Besides, according to Lemma \ref{lem:1}, the final results of the introduced auxiliary variables ${{\bf{T}}_0}$ and $\left\{ {{{\bf{T}}_k}} \right\}_{k = 1}^K$ take the forms, resp.:
\begin{equation}\label{eq:F52}
\begin{array}{l}
{\bf{T}}_0^* = {\left[ {{\bf{I}} + {{\bf{H}}^H}\left( {{\bf{W}}_E^* + {{\bf{V}}^*}} \right){\bf{H}}} \right]^{ - 1}},\\
{\bf{T}}_k^* = {\left[ {{\bf{I}} + {\bf{G}}_k^H\left( {{\bf{W}}_{_I}^* + {{\bf{V}}^*}} \right){{\bf{G}}_k}} \right]^{ - 1}}, \forall k.
\end{array}
\end{equation}
Let ${\phi _k}({{\bf{W}}_I},{{\bf{W}}_E},{\bf{V}}) = {C_I}({{\bf{W}}_I},{{\bf{W}}_E},{\bf{V}}) - {C_{E,k}}({{\bf{W}}_I},
{{\bf{W}}_E},\\{\bf{V}}) - {R_0}\ln 2, \forall k$, substituting (\ref{eq:F52}) into ${x_k}({{\bf{W}}_I}, {{\bf{W}}_E}, {\bf{V}}, \{ {{\bf{T}}_k}\\\} _{k = 0}^K)$, we gain ${\phi _k}({\bf{W}}_{_I}^*,{\bf{W}}_{_E}^*,{{\bf{V}}^*}) = {x_k}({\bf{W}}_{_I}^*, {\bf{W}}_{_E}^*, {{\bf{V}}^*}, [{\bf{I}}+\\{{\bf{H}}^H}({\bf{W}}_E^* + {{\bf{V}}^*}){\bf{H}}]^{ - 1},{\kern 1pt} {\kern 1pt} {\kern 1pt} \{ {[{\bf{I}} + {\bf{G}}_k^H({\bf{W}}_{_I}^* + {{\bf{V}}^*}){{\bf{G}}_k}]^{ - 1}}\} _{k = 1}^K) = x_{_k}^*$. Thus, (\ref{eq:F51}) can be rewritten as
\begin{subequations}\label{eq:F53}
\begin{equation}\label{eq:F53a}
{\nabla _{{{\bf{W}}_I}}}{\rm{E}}_{_{{\rm{WS}}}}^* + \sum\limits_{k = 1}^K {\lambda _k^*\left( t \right){\nabla _{{{\bf{W}}_I}}}\phi _{_k}^*}  - {\upsilon ^*}{\bf{I}} + {\bf{\Phi }}_{{{\bf{W}}_I}}^* = {\bf{0}},
\end{equation}
\begin{equation}\label{eq:F53b}
{\nabla _{{{\bf{W}}_E}}}{\rm{E}}_{_{{\rm{WS}}}}^* + \sum\limits_{k = 1}^K {\lambda _k^*\left( t \right){\nabla _{{{\bf{W}}_E}}}\phi _{_k}^*} - {\upsilon ^*}{\bf{I}} + {\bf{\Phi }}_{{{\bf{W}}_E}}^* = {\bf{0}},
\end{equation}
\begin{equation}\label{eq:F53c}
{\nabla _{\bf{V}}}{\rm{E}}_{_{{\rm{WS}}}}^* + \sum\limits_{k = 1}^K {\lambda _k^*\left( t \right){\nabla _{\bf{V}}}\phi _{_k}^*} - {\upsilon ^*}{\bf{I}} + {\bf{\Phi }}_{\bf{V}}^* = {\bf{0}},
\end{equation}
\begin{equation}\label{eq:F53d}
\lambda _k^*\left( t \right)\phi _{_k}^* = 0, \forall k,
\end{equation}
\end{subequations}
where ${\rm{E}}_{_{{\rm{WS}}}}^*{\rm{ = }}{{\rm{E}}_{{\rm{WS}}}}\left( {{\bf{W}}_{_I}^*, {\bf{W}}_{_E}^*, {{\bf{V}}^*}} \right)$, $\phi _{_k}^* = {\phi _k}\left( {{\bf{W}}_{_I}^*,{\bf{W}}_{_E}^*,{{\bf{V}}^*}} \right)$ and (\ref{eq:F53d}) is result from the previous conclusion $\lambda _k^*\left( t \right)x_k^* = 0, \forall k$. Since $\left( {{\bf{W}}_{_I}^{\rm{*}},{\kern 1pt} {\kern 1pt} {\bf{W}}_{_E}^{\rm{*}},{{\bf{V}}^{\rm{*}}},\left\{ {{\bf{T}}_{_k}^{\rm{*}}} \right\}_{k = 0}^K} \right)$ is the optimal solution of problem $\left( {\overline {\textsc{P3}} } \right)$, it must satisfy the KKT conditions of problem $\left( {\overline {\textsc{P3}} } \right)$, i.e., as listed in (\ref{eq:F54}).
\begin{equation}\label{eq:F54}
\left\{ \begin{array}{l}
\left( {\ref{eq:F51}} \right), {\nabla _{{{\bf{T}}_0}}}\mathcal{L}\left( {\bm{\chi }} \right) = {\bf{0}}, {\nabla _{{{\bf{T}}_k}}}\mathcal{L}\left( {\bm{\chi }} \right) = {\bf{0}}, \forall k,\\
{\upsilon ^*}\left[ {{\rm{Tr}}\left( {{\bf{W}}_{_I}^* + {\bf{W}}_{_E}^* + {{\bf{V}}^*}} \right) - P} \right] = 0, {\upsilon ^*} \ge 0,\\
{\rm{Tr(}}{\bf{\Phi }}_{_{{{\bf{W}}_I}}}^*{\bf{W}}_{_I}^*{\rm{)}} = 0, {\rm{Tr(}}{\bf{\Phi }}_{_{{{\bf{W}}_E}}}^*{\bf{W}}_E^*{\rm{)}} = 0, {\rm{Tr(}}{\bf{\Phi }}_{_{\bf{V}}}^*{{\bf{V}}^*}{\rm{)}} = 0,\\
{\rm{Tr}}({\bf{\Phi }}_{_{{{\bf{T}}_0}}}^*{\bf{T}}_{_0}^*) = 0, {\rm{Tr}}({\bf{\Phi }}_{_{{{\bf{T}}_k}}}^*{\bf{T}}_{_k}^*) = 0, \forall k,\\
{\bf{\Phi }}_{_{{{\bf{W}}_I}}}^* \succeq {\bf{0}}, {\bf{\Phi }}_{_{{{\bf{W}}_E}}}^* \succeq {\bf{0}}, {\bf{\Phi }}_{_{\bf{V}}}^* \succeq {\bf{0}},\\
{\bf{\Phi }}_{_{{{\bf{T}}_0}}}^* \succeq {\bf{0}}, {\bf{\Phi }}_{_{{{\bf{T}}_k}}}^* \succeq {\bf{0}}, \forall k.
\end{array} \right.
\end{equation}
Finally, by combining (\ref{eq:F53}) and parts of the KKT conditions in (\ref{eq:F54}), we note that $\left( {{\bf{W}}_{_I}^{\rm{*}},{\kern 1pt} {\kern 1pt} {\bf{W}}_{_E}^{\rm{*}},{{\bf{V}}^{\rm{*}}}} \right)$  must satisfy the following KKT conditions:
\begin{equation}\label{eq:F55}
\left\{ \begin{array}{l}
{\nabla _{{{\bf{W}}_I}}}{\rm{E}}_{_{{\rm{WS}}}}^* + \sum\limits_{k = 1}^K {\lambda _k^*\left( t \right){\nabla _{{{\bf{W}}_I}}}\phi _{_k}^*}  - {\upsilon ^*}{\bf{I}} + {\bf{\Phi }}_{{{\bf{W}}_I}}^* = {\bf{0}},\\
{\nabla _{{{\bf{W}}_E}}}{\rm{E}}_{_{{\rm{WS}}}}^* + \sum\limits_{k = 1}^K {\lambda _k^*\left( t \right){\nabla _{{{\bf{W}}_E}}}\phi _{_k}^*} - {\upsilon ^*}{\bf{I}} + {\bf{\Phi }}_{{{\bf{W}}_E}}^* = {\bf{0}}, \\
{\nabla _{\bf{V}}}{\rm{E}}_{_{{\rm{WS}}}}^* + \sum\limits_{k = 1}^K {\lambda _k^*\left( t \right){\nabla _{\bf{V}}}\phi _{_k}^*} - {\upsilon ^*}{\bf{I}} + {\bf{\Phi }}_{\bf{V}}^* = {\bf{0}}, \\
\lambda _k^*\left( t \right)\phi _{_k}^* = 0, \lambda _k^*\left( t \right) \ge 0, \forall k,\\
{\upsilon ^*}\left[ {{\rm{Tr}}\left( {{\bf{W}}_{_I}^* + {\bf{W}}_{_E}^* + {{\bf{V}}^*}} \right) - P} \right] = 0, {\upsilon ^*} \ge 0,\\
{\rm{Tr(}}{\bf{\Phi }}_{_{{{\bf{W}}_I}}}^*{\bf{W}}_{_I}^*{\rm{)}} = 0, {\rm{Tr(}}{\bf{\Phi }}_{_{{{\bf{W}}_E}}}^*{\bf{W}}_E^*{\rm{)}} = 0, {\rm{Tr(}}{\bf{\Phi }}_{_{\bf{V}}}^*{{\bf{V}}^*}{\rm{)}} = 0,\\
{\bf{\Phi }}_{_{{{\bf{W}}_I}}}^* \succeq {\bf{0}}, {\bf{\Phi }}_{_{{{\bf{W}}_E}}}^* \succeq {\bf{0}}, {\bf{\Phi }}_{_{\bf{V}}}^* \succeq {\bf{0}},\\
\left( {{\bf{W}}_{_I}^*,{\bf{W}}_{_E}^*,{{\bf{V}}^*}} \right) \in {\bm{{\rm Z}}},
\end{array} \right.
\end{equation}
where ${\bm{{\rm Z}}} = \{ ({{\bf{W}}_I},{{\bf{W}}_E},{\bf{V}})|\;{\phi _k}({{\bf{W}}_I},{{\bf{W}}_E},{\bf{V}}) \ge 0,  {\rm{Tr(}}{{\bf{W}}_I} + {{\bf{W}}_E} + {\bf{V}}{\rm{)}} \le P, {{\bf{W}}_I} \succeq {\bf{0}}, {{\bf{W}}_E} \succeq {\bf{0}}, {\bf{V}} \succeq {\bf{0}}\}$. Clearly, the conditions in (\ref{eq:F55}) are exactly the KKT conditions of problem $\left( {\textsc{P3}} \right)$.



\ifCLASSOPTIONcaptionsoff
  \newpage
\fi

\end{document}